# zCOSMOS: A LARGE VLT/VIMOS REDSHIFT SURVEY COVERING 0 < $z$ < 3 IN THE COSMOS FIELD[1]

S.J. Lilly[2], O. Le Fèvre[3], A. Renzini[4], G. Zamorani[5], M. Scodeggio[6], T. Contini[7], C.M. Carollo[2], G. Hasinger[8], J.-P. Kneib[3], A. Iovino[9], V. Le Brun[3], C. Maier[2], V. Mainieri[8], M. Mignoli[5], J. Silverman[8], L.A.M. Tasca[3], M. Bolzonella[5], A. Bongiorno[5], D. Bottini[6], P. Capak[10], K. Caputi[2], A. Cimatti[11], O. Cucciati[9], E. Daddi[12], R. Feldmann[2], P. Franzetti[6], B. Garilli[6], L. Guzzo[9], O. Ilbert[5], P. Kampczyk[2], K, Kovac[2], F. Lamareille[7], A. Leauthaud[3], J.-F. Le Borgne[7], H. J. McCracken[13], C. Marinoni[9], R. Pello[7], E. Ricciardelli[4], C. Scarlata[2], D. Vergani[6], D. B. Sanders[14], E. Schinnerer[15], N. Scoville[10], Y. Taniguchi[16], S. Arnouts[3], H. Aussel[13], S. Bardelli[5], M. Brusa[8], A. Cappi[5], P. Ciliegi[5], A. Finoguenov[8], S. Foucaud[17], R. Franceschini[4], C. Halliday[11], C. Impey[18], C. Knobel[2], A. Koekemoer[21], J. Kurk[11,15], D. Maccagni[6], S. Maddox[17], B. Marano[19], G. Marconi[20], B. Meneux[6,9], B. Mobasher[21], C. Moreau[3], J.A. Peacock[22], C. Porciani[2], L. Pozzetti[5], R. Scaramella[23], D. Schiminovich[24], P.Shopbell[10], I. Smail[25], D. Thompson[10], L. Tresse[3], G. Vettolani[26], A. Zanichelli[26], E. Zucca[5]

[2] Institute of Astronomy, Department of Physics, ETH Zurich, CH-8093, Switzerland
[3] Laboratoire d'Astrophysique de Marseille, France
[4] Dipartimento di Astronomia, Universita di Padova, Padova, Italy
[5] INAF Osservatorio Astronomico di Bologna, Bologna, Italy
[6] INAF - IASF Milano, Milan, Italy
[7] Laboratoire d'Astrophysique de l'Observatoire Midi-Pyrénées, Toulouse, France
[8] Max Planck Institut für Extraterrestrische Physik, Garching, Germany
[9] INAF Osservatorio Astronomico di Brera, Milan, Italy
[10] California Institute of Technology, Pasadena, USA
[11] INAF Osservatorio Astrofisico di Arcetri, Florence, Italy
[12] National Optical Astronomy Observatories, Tucson, USA
[13] Institut d'Astrophysique de Paris, UMR7095 CNRS, Université Pierre & Marie Curie, 75014 Paris, France and Observatoire de Paris, LERMA, 75014 Paris, France
[14] Institute for Astronomy, University of Hawaii, Honolulu, USA
[15] Max Planck Institut für Astronomie, Heidelberg, Germany
[16] Tohoku University, Tokyo, Japan
[17] Nottingham University, United Kingdom
[18] University of Arizona, Tucson, Arizona, USA
[19] Dipartimento di Astronomia, Universita` degli Studi di Bologna, Bologna, Italy
[20] European Southern Observatory, Garching, Germany
[21] Space Telescope Science Institute, Baltimore, Maryland
[22] University of Edinburgh, United Kingdom
[23] INAF Roma, Italy
[24] Columbia University, New York, USA
[25] Durham University, Durham, United Kingdom
[26] IRA-INAF, Bologna, Italy

[1] Based on observations undertaken at the European Southern Observatory (ESO) Very Large Telescope (VLT) under Large Program 175.A-0839. Also based on observations with the NASA/ESA Hubble Space Telescope, obtained at the Space Telescope Science Institute, operated by AURA Inc., under NASA contract NAS 5-26555, with the Subaru Telescope, operated by the National Astronomical Observatory of Japan, with the telescopes of the National Optical Astronomy Observatory, operated by the Association of Universities for Research in Astronomy, Inc. (AURA) under cooperative agreement with the National Science Foundation, and with the Canada-France-Hawaii Telescope, operated by the National Research Council of Canada, the Centre National de la Recherche Scientifique de France and the University of Hawaii.

ABSTRACT

zCOSMOS is a large redshift survey that is being undertaken in the COSMOS field using 600 hours of observation with the VIMOS spectrograph on the 8-m VLT. The survey is designed to characterise the environments of COSMOS galaxies from the 100 kpc scales of galaxy groups up to the 100 Mpc scale of the cosmic web and to produce diagnostic information on galaxies and active galactic nuclei. The zCOSMOS survey consists of two parts: (a) zCOSMOS-bright, a magnitude-limited I-band $I_{AB} < 22.5$ sample of about 20,000 galaxies with $0.1 < z < 1.2$ covering the whole 1.7 deg$^2$ COSMOS ACS field, for which the survey parameters at $z \sim 0.7$ are designed to be directly comparable to those of the 2dFGRS at $z \sim 0.1$; and (b) zCOSMOS-deep, a survey of approximately 10,000 galaxies selected through colour-selection criteria to have $1.4 < z < 3.0$, within the central 1 deg$^2$. This paper describes the survey design and the construction of the target catalogues, and briefly outlines the observational program and the data pipeline. In the first observing season, spectra of 1303 zCOSMOS–bright targets and of 977 zCOSMOS-deep targets have been obtained. These are briefly analysed to demonstrate the characteristics that may be expected from zCOSMOS, and particularly zCOSMOS-bright, when it is finally completed between 2008-2009. The power of combining spectroscopic and photometric redshifts is demonstrated, especially in correctly identifying the emission line in single-line spectra and in determining which of the less reliable spectroscopic redshifts are correct and which are incorrect. These techniques bring the overall success rate in the zCOSMOS-bright so far to almost 90% and to above 97% in the $0.5 < z < 0.8$ redshift range. Our zCOSMOS-deep spectra demonstrate the power of our selection techniques to isolate high redshift galaxies at $1.4 < z < 3.0$ and of VIMOS to measure their redshifts using ultraviolet absorption lines.

## 1. INTRODUCTION AND MOTIVATION

The overall scientific goals of the COSMOS survey (Scoville et al. 2006a) are to understand the three-way physical inter-relationships between the cosmic evolution of galaxies, their central super-massive blackholes, and the larger scale environment in which they reside. It is expected that the environment, from the 100 kpc scales of groups up to the 100 Mpc scales of the cosmic web of filaments and voids, must be playing a very large and possibly decisive role in the evolution of galactic systems, yet rather little is know at present about the environments of galaxies at high redshift. A related goal is to observe the dark matter distribution directly through the gravitational weak shear of structures along the line of sight to distant galaxies and to relate this to the distribution of galaxies.

The very impressive deep multi-band photometry in the COSMOS field (Taniguchi et al. 2006, Capak et al. 2006) enables the approximate redshifts of vast numbers of galaxies to be estimated from their broad-band spectral energy distributions (Mobasher et al. 2006, Feldmann et al. 2006). At moderate depths $I_{AB} \leq 23$ and $z \leq 1.2$ these photometric-redshifts have a statistical accuracy of $\sigma_z \sim 0.03(1+z)$. This is clearly adequate to identify the regions of highest density (e.g. Scoville et al. 2006b) and to serve as the basis of statistical studies of the galaxy distribution independent of their environment (e.g. Sargent et al. 2006, Scarlata et al. 2006) but is insufficient to delineate the cosmic web and completely inadequate to characterise the environments of galaxies on the scale of galaxy groups – i.e. those environments in which most galaxies actually reside and in which we may expect many of the most important processes that may regulate the evolution and transformation of galaxies to be operating. This is clearly illustrated in Fig 1 where we construct mock surveys at $z \sim 0.7$ using the COSMOS mock catalogues kindly provided by Kitzbichler et al. (private communication) using both spectroscopic and photometric redshifts. The power of spectroscopic redshifts in delineating and characterizing the environments of galaxies therefore motivates a major redshift survey of galaxies in the COSMOS field.

Specific science goals of zCOSMOS cover three broad categories. First, spectroscopic redshifts allow us to generate maps of the large scale structure in the Universe and to quantify the density field throughout the COSMOS volume to $z \sim 3$ with a precision impossible with photometrically estimated redshifts. As well as enabling studies of the variation of galaxy properties with local density, these density maps may be compared with those produced by weak lensing shear maps (Rhodes et al. 2006) and the hot-gas structures detected in X-rays (Finoguenov et al. 2006). The density maps also allow us to determine where in the large scale structure X-ray and radio sources reside and, with absorption line studies of background quasars, to relate the distribution of gas to the large scale distribution of galaxies. More quantitatively, a major goal of zCOSMOS is to generate a catalogue of well characterized groups to determine their number density $N(\sigma)$ and to trace the development of galaxy properties in groups with different physical characteristics, such as crossing time and density, that are likely to be relevant for the evolution of member galaxies. Many statistical measures of the galaxy distribution including the correlation function $\xi(r_p,\pi)$, and the pair-wise velocity dispersion can be determined as functions of galaxy type.

Secondly, the spectra also provide important diagnostics on the galaxies themselves, such as star-formation rates, AGN classification, reddening by dust, stellar population ages and metallicities, as well as metallicities of emission line gas, and the possibility, depending on the velocity resolution, of measuring the internal dynamics of galaxies.

Finally, accurate and reliable spectroscopic redshifts replace and complement photometrically estimated redshifts. Spectroscopic redshifts provide a calibration of photometric-redshift schemes that may then be applied to objects not observed spectroscopically, including those fainter than the spectroscopic limit. By eliminating catastrophic failures, and the sometimes complex redshift likelihood functions for individual objects, spectroscopic redshifts provide a secure determination of the various distribution functions $\phi$(properties) describing the galaxy population. Spectroscopic redshifts and spectral classification also provide confirmation of the identification of X-ray and radio sources.

The main goal of the spectroscopic survey zCOSMOS is thus to characterize galactic environments throughout the COSMOS volume out to redshifts of around $z \sim 3$. At redshifts to $z \sim 1$ it is possible to design a survey that matches very closely the parameters of the very large surveys of the local Universe such as the 2 Degree Field Galaxy Redshift Survey (2dFGRS, Colless et al. 2001), allowing a precise quantitative comparison of structures, and the galaxies within them, over the last 50% of the lifetime of the Universe. At higher redshifts, it is more difficult in practical terms to select galaxies in a directly comparable way, and it is also somewhat harder to measure the redshifts. In particular, some form of colour pre-selection is required to isolate the tail of high redshift galaxies that appears at $I_{AB} > 23$ (see e.g. Le Fèvre et al. 2005).

The VIMOS spectrograph (Le Fèvre et al. 2003) on the 8-m UT3 "Melipal" of the European Southern Observatory's Very large Telescope (ESO VLT) offers a very high multiplexing gain, making a large and densely sampled redshift survey of the large COSMOS field practical. The zCOSMOS redshift survey has been designed to efficiently utilize VIMOS by splitting the survey into two parts. The first, "zCOSMOS-bright", aims to produce a redshift survey of approximately 20,000 *I*-band selected galaxies at redshifts $z \leq 1$ that is directly comparable to the 2dFGRS sample at $z \sim 0.1$ in terms of the sampling rate and redshift measurement success rate, the redshift velocity accuracy and the range of galaxy luminosities covered. Covering the approximately 1.7 $deg^2$ of the COSMOS field (essentially the full ACS-covered area) the transverse dimension at $z \sim 1$ is 75 Mpc. The second part, "zCOSMOS-deep", will observe about 10,000 galaxies selected through well-defined colour selection criteria to mostly lie at $1.5 < z < 3.0$. Simply to keep the required amount of telescope time manageable, the field of zCOSMOS-deep is restricted to the central 1 $deg^2$ of the COSMOS field. However, at $z \sim 2$ the survey subtends a transverse distance of 80 comoving Mpc, slightly larger than the bright part of the survey at lower redshift.

zCOSMOS has been awarded about 600 hours of Service Mode observing on the ESO VLT, making it (Large Program 175.A-0839) the largest single observing project undertaken so far on that facility. Observations started on April 1, 2005, and are expected to take at least three years to complete. These first observations have already allowed us to assess the data quality and predict the ultimate yield of the program.

zCOSMOS, like COSMOS generally, is undertaken in the spirit of a Legacy program, with an emphasis on making the data products of lasting and general usefulness to the broad community of researchers. The purpose of this introductory paper is therefore to explain the motivation for the detailed design of the observational program as it is currently being implemented at the VLT, to describe the construction of the spectroscopic target catalogues and to summarize the observational procedures, as well as the pipeline used to reduce the spectra. We then present the results of various checks undertaken on the first data obtained. These establish at least a preliminary estimate of the reliability of the redshifts, and their velocity accuracy, and allow us to anticipate the properties of the final sample when the observing program is completed. We show how the combination of spectroscopic measurements and photometric redshift estimates can be used to verify the redshifts, break the degeneracies caused by the (relatively few) single line redshifts and identify which of the less reliable spectroscopic redshifts are likely to be correct, further increasing the success rate.

Where necessary, a concordance cosmology with $H_0 = 70$ $kms^{-1}Mpc^{-1}$, $\Omega_{0,m} = 0.3$ and $\Omega_{0,\Lambda} = 0.7$ is adopted. All magnitudes are quoted in the AB system.

## 2. zCOSMOS SURVEY DESIGN

Given the scientific goals described above, the practical design of zCOSMOS is driven by the characteristics of the VIMOS spectrograph. There are a number of trade-offs involving the brightness and number density of the target population and the pattern of telescope pointings, which affect the required exposure time and the success rate in determining redshifts, the total number of objects that are observed and the sampling rate, which we define to be the fraction of targets, selected according to some well-defined criteria, that are actually observed spectroscopically.

The VIMOS spectrograph is a conventional multi-slit imaging spectrograph that can observe simultaneously four quadrants, each roughly $7 \times 8$ arcmin$^2$, separated by a cross-shaped region 2 arcmin wide. The number of slits that may be placed in each mask depends on the length of each spectrum and on the surface density on the sky of the targets. While the total number of objects that may be placed in the masks increases with increasing target density, it is found that the sampling *rate* (i.e. the fraction of available targets for which spectra are obtained) decreases. In designing the masks from an input catalogue, some objects may be designated as "compulsory" targets of special interest, in which case they are included in the mask design if at all possible. The majority of slits are then assigned to "random" targets in the catalogue, selected so as to maximise the number of slits in each mask. Some objects may of course also be "forbidden" (e.g. if previously observed) and not included. It is found statistically that the addition of each compulsory target reduces the number of random slits by two.

The quadrant design of VIMOS means that a large contiguous area can be covered uniformly by stepping the field centers, or "pointings", across the larger survey field by an amount in each direction that is equal to the dimensions of the individual VIMOS quadrants. We define the "coverage" as the number of opportunities that a given point on the sky has to be included in a mask and thus observed spectroscopically. The "sampling rate" is then the fraction of real targets that are observed in the masks, which will depend on the coverage and the local density of the targets. This uniform pattern of pointings (see Fig 2) produces a large contiguous region of coverage equal to four, surrounded by four edge regions (of width the quadrant size plus the gap) with coverage two, and four small corner regions with coverage equal to one. Repeating the pattern of pointings by designing more than one mask for each pointing doubles these coverages. The multiple-pass strategy, which is mandatory to achieve uniform coverage, has the considerable benefit of producing a high coverage, substantially reducing the bias against near neighbours that is otherwise inherent in slit spectrographs (see Fig 4 below).

The primary science goal of zCOSMOS is to trace the large scale structure in the Universe at high redshifts and to characterize both linear and non-linear density enhancements such as galaxy groups. Ideally, the design goals were to ensure:

- a high and uniform sampling rate across the field, with a goal of 70%;
- a high success rate in redshift determination, defined as the fraction of objects actually observed that ultimately yield a reliable redshift, with a goal of 90%.
- velocity accuracies of order 100 kms$^{-1}$ enabling dynamical characterisation of the environment down to low mass scales.
- a more or less contiguous redshift coverage in the COSMOS field over $0 < z < 3$, spanning 85% of cosmic time and containing the peak in the global star-formation rate and AGN activity.

Simulations with the mask design software (Bottini et al. 2005) indicated that sampling rates of approximately 70% can be achieved with VIMOS for a target density of about 20,000 $deg^{-2}$ with the four-pass strategy described above for the LR-Blue grism or with an eight-pass coverage (with two mask designs at each pointing) for the longer spectra that are produced by the MR grism and OS-Red filter. Taking into account these scientific and technical considerations, there is thus an optimal survey configuration in two regimes, each having an input catalogue with about 20,000 galaxies $deg^{-2}$.

## 2.1 zCOSMOS-bright

The brighter, lower redshift component of zCOSMOS has a pure magnitude selection at $I_{AB}$ < 22.5 as used in the CFRS (Lilly et al. 1995a) and VVDS-wide surveys (Le Fèvre et al. 2005). This selection yields redshifts in the range $0.1 < z < 1.2$. The velocity accuracy of below 100 $kms^{-1}$ requires the R ~ 600 MR grism, necessitating an 8-pass sampling strategy and one hour integrations to secure redshifts with a high success rate. The spectral range is in the red (5550 – 9650 Å) to follow the strong spectral features around 4000 Å to as high redshifts as possible. This observational set up yields a sample that is directly comparable with the low redshift 2dFGRS at $z \sim 0.1$ in terms of selection, in sampling and success rates, and in velocity accuracy, as described in more detail below.

## 2.2 zCOSMOS-deep:

In order to isolate galaxies at $1.5 < z < 2.5$, from the much larger number of low luminosity galaxies at z < 1 (see Le Fèvre et al. 2005), some kind of colour selection must be applied. The use of well-defined color criteria for spectroscopic target selection is to be preferred over using the output of a photometric redshift scheme, to ensure that the sample is uniquely and repeatably defined.

At least two methods have been demonstrated to be effective at isolating such galaxies: the BzK criteria of Daddi et al. (2004) and the ultraviolet UGR "BX" and "BM" selection of Steidel et al. (2004) which merges into the well-known U-dropout selection (Steidel et al. 1996) at higher redshifts.

The BzK criterion has the advantage of selecting both actively star forming as well as passively-evolving galaxies in the range $1.5 < z < 2.5$. Moreover, the star-forming BzK-selected galaxies are on average more massive, more dust-obscured, and have higher star-formation rates than UGR-selected galaxies (Reddy et al. 2005, 2006). To the limit $K_{AB}$ < 21.8, over 80% of them are detected at 24 μm with Spitzer/MIPS in the GOODS field (Daddi et al. 2005, Reddy et al 2005) and many qualify as Ultraluminous Infrared Galaxies (ULIRGs). However, their surface density to the limit of feasible optical spectroscopy and current *K* limits (about $10^3$ $deg^{-2}$, Kong et al. 2005), is too low to trace the LSS with the desired accuracy, and for an optimal exploitation of the multiplex of VIMOS. Thus, only combining the UGR and the BzK criteria one can ensure a fairly complete inventory of star-forming galaxies, trace the LSS to the required detail, and fully exploit the capabilities of the spectrograph. In the observations carried out in 2005, the BzK-selection has been limited to $K_{AB}$ < 21.85 (see below). In the meantime considerably deeper K-data in the COSMOS field has been obtained as well as Spitzer 3.6 and 4.5 μm data. These should allow a more optimised selection in the future.

Galaxies in this redshift range are best observed with VIMOS in the blue spectral region to pick up the stronger absorption features in the range between 1200 Å and 1700 Å. This effectively requires that we apply an additional magnitude selection $B_{AB} \leq 25.0$ to ensure an adequate signal in the continuum. This eliminates most of the passively-evolving galaxies at $z \sim 2$. These galaxies would be best observed with the LR-Red grism, but their surface density is too low for an efficient use of VIMOS and will have to be observed at other facilities. Clearly, the census of galaxies at $1.5 < z < 2.5$ will not be complete until redshifts are secured also for this kind of galaxies. These are likely to be the most massive in this redshift range, inhabiting the highest-density peaks in the large scale structure of the cosmic web. To complete the redshift coverage by including the passively-evolving BzKs remains as a major goal of the broader COSMOS project.

Measurement of secure redshifts in this redshift range down to $B_{AB} \sim 25$ requires 4 to 5 hours of integration with the R ~ 200 LR-Blue grism, which gives a spectral range from 3600 – 6800 Å. Compared to zCOSMOS-bright, this set up extends the survey through the relatively unexplored "redshift desert" to $z \sim 3$, albeit with a less straightforward selection function, somewhat lower velocity accuracy and, it is expected, a slightly lower success rate in measuring secure redshifts.

## 3. COSNTRUCTION OF THE INPUT CATALOGUES

### 3.1 The "zCOSMOS-bright" Galaxy Catalogue

At the relatively bright magnitudes of zCOSMOS-bright, it is possible to construct a catalogue with selection criteria that are well matched to those that have been used to construct the very large spectroscopic surveys, e.g. 2dFGRS and the SDSS, in the local Universe, facilitating direct comparisons over more than a half the Hubble time.

There are complementary advantages and disadvantages to using either HST or ground-based images alone for the catalogue generation and the optimum strategy is to combine both approaches. The primary input catalogue was generated using SExtractor (Bertin et al. 1996) applied to the COSMOS F814W HST/ACS images sampled at 0.03 arcsec pixel$^{-1}$ (Koekemoer et al. 2006, Leauthaud et al. 2006) in a "hot and cold" two-pass process to first identify bright objects. This substantially reduced the tendency of the HST-based catalogue to "over-resolve" extended galaxies into multiple components. This initial SExtractor catalogue was then "cleaned" by carrying out a detailed comparison with one extracted from a stack of *i** images obtained with MEGACAM on the 3.6m Canada-France-Hawaii telescope and processed at the TERAPIX data reduction centre in Paris. There is no significant systematic photometric offset between these two catalogues at $I_{AB} \sim 22.5$ but naturally there is some scatter which causes objects to cross the boundary between the samples. Objects with magnitudes within 0.3 mag were not considered significantly discrepant and primacy in these cases was given to the ACS-based magnitudes.

All the more significant discrepancies were visually examined over the entire field and resolved by eyeball inspection of the images. These discrepancies had a number of origins that could be easily dealt with:

- objects missing because they had been masked out of one or other catalogue due to bright stars. Fortunately, the diffraction patterns in the two sets of images were rotated by about 10 degrees, enabling the CFHT image to cover the extended diffraction spikes in the HST images, making the final inaccessible area very small. Objects missing in the ACS catalogue were inserted with their CFHT magnitudes.
- objects lying just outside of the jagged boundary of the ACS imaging and in two small areas not observed by ACS. These were also brought into the catalogue with their CFHT magnitudes.
- multiple objects that had been blended together by CFHT. Unless it was clear that in fact they represented a single galaxy incorrectly broken up into individual HII regions, by the superior resolution of the HST, these multiple objects were retained, provided that they were individually brighter than $I_{AB} = 22.5$ in the ACS catalogue. Objects which were individually fainter than 22.5 in the ACS catalogue were excluded, even if the CFHT had lumped them together to make an object above the $I_{AB}$ = 22.5 threshold, unless it was clear that this was in fact a single object "over-resolved" by HST/ACS. The importance of this is that the catalogue should generally represent close pairs of galaxies as two galaxies at their correct magnitudes rather than as a single brighter galaxy – thereby eliminating one of the major biases potentially present in ground-based samples.

- a very small but non-zero number of objects that were missing in the ACS catalogue for no obvious reason that were ascribed to SExtractor "failures" and replaced with the CFHT photometry. A correspondingly small number of ACS detections were rejected as spurious if there was no sign at all of a suitably bright object on the CFHT image.

The resulting catalogue was compared with a similar but independent catalogue that had been produced early in 2005 using the same procedure applied to the Cycle 12 ACS images sampled at 0.05 arcsec pixel$^{-1}$, with again a second exhaustive visual check of discrepancies. This lead to the re-insertion of an additional 50 or so objects, representing 0.1% of the final catalogue. It should be noted that this earlier catalogue was used for the 2005 spectroscopic observations and because of small differences in the photometry, both systematic at the 0.01 mag level and random at the few hundredths of a magnitude level, about 4% of the objects in the first catalogue do not appear in the full-field second catalogue which will be used for the remainder of the spectroscopic program. These objects will therefore appear in a supplementary catalogue and we will be able to present a statistical weighting based on the number of masks so that the correct statistical weighting can be applied.

Each object in the catalogue is represented as a single potential spectroscopic target in the mask designing process, unless two or more targets (both with $I_{AB} < 22.5$) were located within 0.6 arcsec in which case they were regarded for practical purposes as a single spectroscopic target with a record retained of the fact that the target was in fact "multiple". Such targets merit careful scrutiny of the spectrograms.

As a last step, the final catalogue was compared with the independent COSMOS catalogue (Capak et al. 2006) derived from very deep Subaru I-band imaging. Spot checks of random areas indicated that any remaining problems of missing objects involve less than 1% of objects.

The procedure above was designed to ensure that the final input catalogue for the zCOSMOS-bright survey represents the cleanest possible sample that may be compared with local samples. The zCOSMOS-bright catalogue is intended to be simply defined as having an ACS/HST SExtractor "magauto" brightness in the range $15.00 < I_{AB}(814) < 22.50$.

This catalogue contains 52,792 objects in 1.91 deg$^2$. Of these, 10,205 objects (19.4%) are classified as stars in that they are both unresolved on the HST images *and* have a UBVRIZK spectral energy distribution that is better matched by a stellar template than by either a galaxy or quasar template at any redshift. These objects were excluded from the spectroscopic target list bringing the final target density to about 22,300 deg$^{-2}$, which is close to the optimal level discussed above. In fact, 5% of the objects observed spectroscopically are identified as stars at zero redshift, suggesting that the exclusion of stars has probably erred on the side of caution. The final number counts of stars and galaxies are shown in Fig 3 and demonstrate excellent agreement with the wide field galaxy counts of Postman et al. (1998).

### *3.2 The High Redshift zCOSMOS-deep Galaxy Catalogue*

Deep imaging of the COSMOS field at longer wavelengths is still being obtained. Of particular importance for the selection of high redshift objects will be the ground-based imaging at K and the Spitzer/IRAC imaging at 3.8 and 4.5 μm. The final target catalogues for zCOSMOS-deep will not be finalized until the Fall of 2006 in preparation for the 2007 observing season.

For the spectroscopic observations undertaken in 2005, a preliminary catalogue was generated from an early (February 2005) version of the COSMOS deep multicolor photometric catalogue (Capak et al. 2006) as an $I_{AB} < 25$ selected sample. At that time there was no G-band data, necessitating a modification of the Steidel et al. (2004) UGR selection criteria. A simple selection in the (U-B)/(V-R) color-color plane was adopted. Relative to Steidel's (U-G)/(G-R) criteria, this had the advantage that photometric errors in the two colors are uncorrelated, but the disadvantage that objects with a continuum break between B and V could satisfy the selection criteria, possibly leading to an increase in the contamination of the high redshift sample by very low redshift interlopers. In the case of the BzK selection (Daddi et al. 2004), we were able to use the normal selection criteria, although the limited depth of the near infrared COSMOS images in 2005 required us to impose a relatively bright magnitude limit of $K_{AB} < 21.85$, approximately the 5$\sigma$ limit of the photometry (Capak et al 2006). For both the "UGR" and "BzK" selections, we then applied an additional selection of $22.5 < B_{AB} < 25.0$ in order to ensure sufficient continuum to detect the ultraviolet absorption features. Straightforward 3 arcsec aperture magnitudes were used to define both the colours and magnitudes.

For the bulk of the observations for zCOSMOS-deep, which will begin in earnest in the 2007 observing season, we will utilize new G-band photometry in the COSMOS field (Capak et al 2006) and deeper K-band imaging over the central square degree that now reaches down to $K_{AB} \sim 23.5$. In addition, we will explore the advantages of using the new Spitzer/IRAC photometry (Sanders et al 2006) which reaches to AB(3.6$\mu$m) ~ 23.5.

### *3.3 X-ray, Radio and Ultraviolet Selected Targets*

Target lists of X-ray (Hasinger et al. 2006, Brusa et al. 2006) and radio identifications (Schinnerer et al. 2006), plus a limited number of Galex-selected targets (Schiminovich et al., in preparation) are included in the mask design as both "compulsory targets" and as "random targets" (see Section 2 above), regardless of whether they satisfy the selection criteria of the main zCOSMOS-bright and zCOSMOS-deep samples. These additional targets are inserted into the bright masks if they have $I_{AB} < 23.5$, otherwise they were observed in the zCOSMOS-deep masks. As random targets, these identifications are observed with the same probability as the normal galaxy targets, i.e. approximately 70%. The compulsory targets are included in the masks at a higher rate. It should be noted that many of these additional targets also satisfy the selection criteria for the main redshift survey. This is clearly inconsequential for those included as random targets, but it should be appreciated that the statistical treatment of these objects within the main survey will require special care because they will have been included with a higher sampling rate than objects that were not so identified. It is expected that of order 700 X-ray sources (Hasinger et al. 2006) will eventually be included as compulsory targets. Approximately 1000 radio source identifications (Schinnerer et al. 2006) should be observed in zCOSMOS-bright, with an additional 1000 in the zCOSMOS-deep program.

## 4. SPECTROSCOPIC DATA AND PIPELINE PROCESSING

### 4.1 Mask Design

The mask design, i.e. the placement of slits over target galaxies, was undertaken with the VMMPS software (Bottini et al. 2005), using the input catalogues described above and 180 second R-band "pre-images" that had been obtained previously in Service Mode with VIMOS in imaging mode. The minimum slit length was set at 10 arcsec. The software works across the field, placing a slit over the next available object (i.e. the first centered more than 5 arcsec from the end of the previous slit). Additional slits are, where possible, added above or below, with the same slit length, at sufficient distance that the first-order spectra do not overlap. The slits are aligned east-west to minimize slit losses from atmospheric differential refraction as the zenith angle increases, and have a width of 1.0 arcsec. In order to maximise the number of slits on the mask. The vast majority of spectra extend over the full wavelength range. However, in designing the slit mask, up to 500 Å are permitted to be lost due to the ends of the spectra falling off the detector area.

Typically, our experience so far has been that, with the current input catalogues, about 160 slits can be placed in the four-quadrant mask for the R ~ 600 MR-Red grism, and about 250 slits in each four-quadrant mask for the R ~ 200 LR-Blue grism on the first two passes. On subsequent passes the numbers reduce somewhat due to the fact that the number of available targets is lower.

At the time of writing, enough masks have been designed for the 8-pass zCOSMOS-bright program that we may assess, from a small but representative area in which all eight masks have been designed, the actual final sampling rate that is achieved in practice. This is found to be 67%, the small reduction from the designed 70% being due to the inclusion of a small percentage of compulsory targets. Fig. 4 shows the sampling rate as a function of pair separation, i.e. of the distance from another object observed. The sampling rate is more or less independent of separation above 10 arcsec and shows only a small reduction at smaller separations, staying always above 50% even for very close pairs. The average sampling rate corresponds to 15,100 observed objects deg$^{-2}$ which will yield a projected final sample of approximately 20,000 galaxies observed in the zCOSMOS-bright sample taking into account the four regimes of coverage factor.

## 4.2 Observations

Observations for zCOSMOS are executed at the VLT in Service Mode. For the brighter MR-Red observations, one hour of integration is obtained in five 720 second exposures, between which the telescope was offset in a pattern of positions separated by one arcsec along the slit. Observations with the LR-Blue grism are split into three "Observation Blocks", executed independently, each of which consists of five 1080 second integrations with the same spatial offsetting, yielding a total integration time of 16,200 seconds (4.5 hours). Because of the lack of an atmospheric dispersion corrector on VIMOS, observations in the blue (zCOSMOS-deep) are limited to ± 2 hours of Hour Angle, and those in the red (zCOSMOS-bright) to ± 2.5 hours of Hour Angle. All observations were executed during periods when the seeing was better than 1.2 arcsec.

## 4.3 Data Reduction

End-to-end data reduction is carried out using the VIPGI software package (Scodeggio et al. 2005). The data reduction is quite automated. Nevertheless, there are still many points of human intervention, especially in the crucial area of redshift determination, and it is advantageous to undertake two completely independent end-to-end reductions of the data at two different institutes, followed by a "reconciliation" between the final results.

VIPGI undertakes standard processing of bias subtraction and flux and wavelength calibration of the spectra, identification of objects in the slit profile and extraction of the one-dimensional spectra. Wavelength calibration uses a HeNeAr arc lamp exposure obtained immediately following the science exposures. At present no flat-fielding correction is carried out.

Determination of redshifts is a multi-step process and involves the use of different approaches tailored to the individual spectra. These include first a fully computer-aided determination based on cross-correlation with template spectra coupled to continuum fitting and principal-component analysis, using the KBRED software, which are a set of routines implemented in the IDL environment (Scaramella et al., in preparation). The success of the automatic measurements is dependent on the use of a representative set of galaxy templates, which we have drawn from those built for the VVDS (Le Fèvre et al. 2005) and from the current program. This preliminary automated step is followed by a detailed visual examination of the one- and two-dimensional spectrograms of every object to assess the validity of the automated redshift. In cases where the automatic procedure fails, a new redshift is computed based on the wavelengths of recognized features. These various automatic and manual tools are smoothly embedded within VIPGI, allowing a quick cross-comparison of the results from the different methods over the actual spectrum.

This process concludes with the assignment of both a redshift and a confidence class to each spectrum. This measurement is then reviewed by a second independent person, followed by a face-to-face "reconciliation" of the result with that from the independent reduction in the second institute. Only then is a final redshift and confidence class assigned. It is important to stress that this procedure means that every zCOSMOS spectrum is examined visually on at least five different occasions by a minimum of four different individuals. This duplication is necessary to assure the highest possible quality control.

Redshift measurements in any redshift survey span in practice a range of reliability from completely secure to a small number of objects for which the redshift is highly unreliable. Different scientific applications may have differing tolerances to incorrect redshifts etc., and characterization of the confidence associated with each redshift is therefore very important. The zCOSMOS Confidence Classes are similar in spirit to those adopted in the CFRS (Le Fèvre et al. 1995) and VVDS (Le Fèvre et al. 2005). It is important to note that they are based on the confidence in the redshift and not on the quality of the spectrum. To briefly summarize:

Class 4 — A completely secure redshift based on unambiguous multiple spectral features of the expected relative strength which leave no room for any doubt at all about the redshift.

Class 3: A very secure redshift, but one for which the classifier(s) recognises at least a remote possibility for error, e.g. because supporting features are in a noisy part of the spectrum, an expected absent feature is similarly noisy and so on. The distinction between this Class and Class 4 is not very well defined and for many purposes they may be combined.

Class 2: A less secure redshift, for which the claimed redshift is by far the most likely, but for which a significant possibility remains that the redshift is nevertheless incorrect.

Class 1: A possible redshift, for which there is a substantial chance that the redshift is in fact wrong, i.e. virtually an informed guess.

Class 0: Cases where no redshift could be ascertained.

Class 9: These are a special case involving a single narrow strong isolated emission line of undoubted reality. Generally, this can only be H$\alpha$ 6563, [OII] 3727 or, at very high redshifts, Ly$\alpha$. For such an object, Class 9 is assigned if no secure choice between these possibilities can be made, choosing the identification that appears more probable, i.e. least unsupported since the Class 9 objects are such because of the absence of supporting spectral features. In practical terms, Class 9 objects in the bright LR-Red spectra could be either at $0.75 < z < 1.5$ (for [OII] 3727) or at $z < 0.45$ (for H$\alpha$). We show below that photometric redshifts in fact confirm the identification for the vast majority of Class 9 spectra and allow the ambiguity to be removed for the remainder.

Class +10: Adding a 10 to the Class signifies a broad line AGN.

The distribution of Confidence Classes within the survey, and an empirical check of the reliability of redshifts in the different classes is presented below in Section 5.2 and Table 2. We show in Section 5.3 that very consistent results are obtained when comparing with photometric redshift estimates and that these can be used to resolve the degeneracies inherent in the Class 9 redshifts, and to identify which of the less reliable redshifts are likely to be incorrect.

## 5. FIRST SEASON OBSERVATIONS OF zCOSMOS-bright IN 2005

The first zCOSMOS-bright spectroscopic observations were carried out in VLT Service Mode during the period April to June 2005. Eight masks were observed, one of which was repeated, yielding spectra of 1303 objects. In terms of masks this is just 4% of the complete program, in terms of spectra, about 6%. This initial quite limited set of data enables us to assess (a) the quality of the data and, for the repeated MR-Red observations, obtain an estimate of the redshift accuracy and reliability of each Confidence Class, and (b) the overall redshift distribution and efficacy of the colour selection criteria, especially for the faint sample. In this paper we consider below the statistics we have derived from these 2005 observations. Initial scientific results from these observations will be presented elsewhere.

The VIMOS MR spectra are of high quality. Fig 5 shows coadded spectra of emission and absorption line galaxies, showing the clear separation of H$\alpha$ and [NII] 6583. In individual spectra it is found that we can typically measure emission lines with an equivalent width of 4-5 Å or 2-3$\times10^{-20}$ Wm$^{-2}$ with an accuracy of about 10%.

### 5.1 Redshift accuracy

One LR-Red mask (55B) was observed twice, separated by several days. Three quadrants with two observations were reduced completely independently (at a different pair of institutes), and underwent two independent "reconciliations" in an identical manner to that employed for the other data. Only at end were these two streams compared. These spectra therefore allow a first assessment of the measurement accuracy of the zCOSMOS MR redshifts, including effects arising from (a) non-repeatable mask positioning within the spectrograph; (b) the data reduction process including wavelength calibration; and (c) the process of redshift measurement.

Of the 116 objects in these three quadrants, 90 were assigned the same redshift on both occasions and therefore provide an empirical estimate of the redshift accuracy (note that the number of discrepancies is of course roughly double that expected in a single observation). Fig. 6 shows the velocity differences of these 90 measurements. The distribution is quite well characterised by a Gaussian of standard deviation 78 kms$^{-1}$, from which we infer that the r.m.s. velocity accuracy of each measurement is of order 55 kms$^{-1}$. This is well within the design error budget of 100 kms$^{-1}$.

One remaining potential source of velocity error comes from positioning of the objects relative to the 1 arcsec slits, especially if this is a function of location within the spectrograph field. We will be able to quantify this component later in the program when observations will be repeated with quite different telescope pointings.

### 5.2 Spectroscopic Verification of Reliability of Confidence Classes

The 116 objects observed twice (see Section 5.1) provide 232 opportunities to check the reliability of redshifts in the different Confidence Classes. The distribution of Confidence Classes amongst these 232 classifications is very similar to that amongst the zCOSMOS-bright sample observed so far. We regard a valid "test" to be one where the other observation (not distinguishing the temporal order) was assigned an equal or higher Confidence Class in the sequence 1 - 2 – (3 and 9) - 4. In other words, two Class 3 observations provide a test of both, whereas a Class 4 compared with a Class 2 provides a test of the 2 but not of the 4. This gives 174 valid tests.

Table 2 summarizes these comparisons: The success rate of Class 4 and Class 3 is currently 100% (111/111 and 34/34 respectively). Of Class 9, the redshift agrees in 85% of cases and in 100% (12/12) once the two-fold degeneracy between a high and low redshift is recognised. Class 2 redshifts are verified in 85% of the tests (11/13), whereas the Class 1 redshifts are, as expected, only confirmed 50% of the time. It should be noted that the 100% reliable Class 4, 3 and 9 comprise 79% of all spectra, with an additional 7% in the 85% reliable Class 2. Class 1 and Class 0 comprise 14% of the spectra.

### 5.3 Use of Photometric Redshifts to Complement the Spectroscopy

The high quality of the photometry in the COSMOS field and of the derived photometric redshifts at $z < 1.2$ suggests that the photometric redshifts may be used to further assess the reliability of the spectroscopic redshifts in the zCOSMOS-bright sample. The comparison between spectroscopic and photometric redshifts is shown in Fig. 7. We accept consistency between photometric and spectroscopic redshifts if the redshift difference $|z_p - z_s|$ is less than $0.1\times(1+z)$, i.e. approximately three times the standard deviation of the differences achieved in the population. We look first at the simple case of the single-line redshifts.

#### 5.3.1 Identification of Single Line Redshifts

As noted above, the spectra of about 7% of galaxies objects show only a single emission line that cannot be reliably identified because of the absence, or weakness, of supporting features. For the zCOSMOS-bright sample, this line is most likely to be either Hα 6563 at $z < 0.5$ or [OII] 3727 at $z > 0.5$. An identification with Hβ 4861 or [OIII] 4959, 5007 can usually be ruled out from the absence of the other lines unless the spectrum in this region is badly affected by telluric emission or other defects. Usually it is possible to prefer one redshift over the other, and in the pipeline processing these spectra are assigned a Confidence Class of 9 at that redshift, while recognising that at least one other redshift is also a possibility.

The upper right panel in Fig. 7 plots the photometric redshift against the proposed (Class 9) spectroscopic redshift for these sources, together with the loci expected if the line has been misidentified. This figure strongly suggests that 65 of the 77 (85%) of the Class 9 redshifts were in fact already correctly identified. This is in excellent agreement with the spectroscopic analysis on 12 objects discussed in Section 5.2. For seven of the Class 9 objects (8%) the line was tentatively identified as [OII] 3727 but the photometric redshift strongly suggests that it is in fact Hα at lower redshift. For the remaining five objects, the comparison is inconclusive, although we note that three of them in any case have poorly fitting templates in the photometric redshifts, perhaps because of the strong emission line itself.

#### 5.3.2 Verification of Spectroscopic Confidence Classes

We may extend this approach to the other objects. The remaining panels of Fig. 7 compare the photometric redshifts of the objects in the first set of spectroscopic redshifts, as a function of the spectroscopic Confidence Class, distinguishing between those where the template fit in the photometric redshift code that we have used (Feldmann et al. 2006) is good and poor (the latter photometric redshifts are less reliable). As would be expected, the agreement is excellent for the secure Class 3 and 4 redshifts, and poorest for the Class 1 spectroscopic redshifts, which as noted above are only expected to be correct 50% of the time. Table 2 summarizes the statistics of agreements and disagreements as a function of the Confidence Class.

There is again a rather striking agreement with the verification rates derived above from the spectroscopy alone for the small number of repeat observations. This suggests we may use the photometric redshifts to determine which of the less reliable spectroscopic redshifts are actually correct, further increasing the spectroscopic success rate to almost 90% while keeping the number of "interlopers", i.e. objects which have an incorrect spectroscopic redshift, in the sample very small. We intend to incorporate such a modified classification scheme to include the information on the consistency with the photometric redshift in future releases of the data. We may also use the photometric redshifts to return to those spectra which appear to be discrepant and search for a redshift in a more limited range, although we intend to always produce a first redshift estimate independent of the photometric redshift.

### 5.3.3 Objects without Secure Redshifts?

Finally, we can ask what the photometric redshifts and spectral types for the objects that we fail to secure a redshift for, considering both the objects in Confidence Class 0 and the 15% of objects in the Confidence Class 1 for which the photometric redshift is substantially discrepant from the proposed spectroscopic redshift.

In the left-hand panel of Fig. 8, we plot the distribution in color-redshift space of the 89% of the current sample for which we believe we have successfully measured a redshift, i.e. either with a very secure Class 3 or 4 redshift or with a less secure Class 2 or 1 redshift that is nevertheless consistent with the photometric redshift. The right hand panel shows the distribution of the remaining 11% of "failures", i.e. Class 0 and Class 1 and 2 with an inconsistent photometric redshift, using the photometrically estimated redshift. The visible vertical banding in the left hand panel of Fig 8. is real and demonstrates the existence of large density variations within the field (see also Fig. 10).

It is clear that the failure rate is a strong function of redshift. This is shown in Fig. 9. The success rate shows a broad maximum around $z \sim 0.7$, and between $0.5 < z < 0.8$ over 97% of galaxies appear to have a successful redshift measurement with our observational set up. The success rate then falls off both to high redshifts $z \sim 1$ and at lower redshift $z < 0.3$.

## 5.4 Redshift Structure in the COSMOS Field

Fig. 10 shows the redshift distribution of objects in this initial set of data. As expected it is highly structured reflecting the existence of large scale structure in the COSMOS field. This emphasizes that COSMOS is not immune to so-called sampling variance, although in this context it should be noted that the objects shown in Fig. 10 are selected from about 20% of the overall COSMOS field (see Fig. 11).

# 6. FIRST SEASON OBSERVATIONS OF zCOSMOS-DEEP IN 2005

Four spectroscopic masks were observed for zCOSMOS-deep in the 2005 observing season. These yielded 977 spectra which were processed, in duplicate, as described above for the bright spectra. Direct verification of the reliability of the different Confidence Classes through repeat observations has not yet been possible, but the reliability of the classes should be similar to those in zCOSMOS-bright described above.

These first observations already demonstrate that the lower resolution of the LR-Blue grism is adequate to detect the ultraviolet absorption lines in the spectra of faint high redshift galaxies.

This is demonstrated in Fig. 12 which shows some composite spectra of high redshift galaxies from zCOSMOS-deep.

Figure 13 (lower right panel) shows the overall redshift distribution of these 977 spectra. There is a small stellar component (< 3%), and a low redshift "interloper" contamination (generally at $z < 0.3$) of 15%. A half of the latter are relatively bright ($B_{AB} < 24$) and several of them are actually now outside of the nominal selection criteria with new photometry. In order to directly compare the redshift distributions with published work, we have used the latest COSMOS photometry (new G-band magnitudes from Capak et al 2006 and deeper unpublished K-band data) to extract subsets of the sample of targets that was actually observed in 2005 which today, with the new photometry, still satisfy the UGR BX and BM criteria (Steidel et al 2004) or the BzK criterion (Daddi et al 2004). For the latter, we take objects with $K_{AB} < 23.5$, and for the former, we apply color corrections to the COSMOS photometry as follows:

$$(U\text{-}G)_{STEIDEL} = 1.19\,(U-G)_{COSMOS}$$
$$(G\text{-}R)_{STEIDEL} = (G-0.5R-0.5I)_{COSMOS}$$

The redshift distributions of these subsets are shown in the remaining three panels of Figure 13.

It is clear that the success rate of redshift determination varies with the selection criteria, and thus with redshift. It is highest for the UGR-BX sample, where 75% of spectra yield a redshift with Confidence Class ≥ 2 (only a little worse than zCOSMOS-bright) and lowest for the UGR-BM sample where this fraction falls to 45%. Most (about 75%) of our targets have $24 < R < 25$ and our overall success/interloper rates for our objects in this magnitude range are 72% / 3% for BX and 45% / 4% for BM. These are broadly comparable to the equivalent rates given by Steidel (2004) which are 65% / 5% and 58% / 4%.

In zCOSMOS, there is a bigger difference between BX and BM. This undoubtedly reflects the fact that redshifts are easier to measure at $z > 1.9$ than at lower redshifts because of the presence at the higher redshifts of the strong features between 1200 – 1600 Å (i.e. CIV 1549, SiIV+OIV 1399 and Ly $\alpha$ 1216) in our spectral range. Based on the redshift distributions in Figure 13 and the expected redshift distributions for BX and BM objects (Steidel et al (2004) indicated by the horizontal bars in the relevant panels, we suspect that most of the objects for which we could not determine a redshift (Class 0) actually lie between $1.4 < z < 1.9$. We also suspect that when we have made a mistake in assigning a redshift for the lower Confidence Classes (i.e. as is presumably the case for ~ 50% of Class 1 and ~ 15% of Class 2 spectra), then the sense of this mistake has been to assign a $z > 1.9$ to an object that in reality lies at $z < 1.9$.

As zCOSMOS-deep embarks on the major part of the program in 2007 and later, we may decide to eliminate the $z < 1.8$ redshift range from the target list to maximise the number of secure redshifts that are obtained, and to extend the redshift range beyond $z > 2.5$, where redshifts should be correspondingly easy to measure.

## 7. COMPARISONS WITH OTHER SURVEYS

In this section we offer brief comparisons with the anticipated final zCOSMOS samples with three relevant comparison samples, the 2dFGRS at low redshift, the $I_{AB} < 24$ selected VVDS-wide and -deep surveys, and the color-selected DEEP2 sample.

### 7.1 Comparison between 2dFGRS at $z \sim 0.1$ and zCOSMOS-bright at $z \sim 0.75$

The 2dFGRS (Colless et al. 2001) is a large local redshift survey of 221,414 galaxies with $b_J < 19.45$. The low redshift *b*-selection of 2dFGRS is broadly equivalent to the I-selection of zCOSMOS-bright at redshifts $0.5 < z < 1.0$ and as shown in Fig. 14, zCOSMOS galaxies in this range of redshifts have luminosities (removing an assumed luminosity evolution $\Delta M_B \sim z$) of $M_B < -19$. Thus the $0.5 < z < 1.0$ zCOSMOS-bright sample is particularly well matched to the 2dFGRS at redshifts $0.076 < z < 0.16$. The redshift accuracy of zCOSMOS is comparable or slightly better than the 85 $\mathrm{kms}^{-1}$ of the 2dFGRS while the product of the sampling and success rates is only slightly worse, 63% vs. 84%. In almost all regards, the 2dFGRS sample in the range $0.076 < z < 0.16$, thus represents an ideal comparison sample with the higher redshift $0.5 < z < 1.0$ regime of zCOSMOS-bright. Not surprisingly, the zCOSMOS-bright sample, observed at half the Hubble time, will be an order of magnitude smaller than the present-epoch 2dFGRS.

Given the similarities of these two samples, we may for instance use the 2PIGG group catalogue (Eke et al. 2004) to anticipate the properties of a group catalogue constructed from zCOSMOS-bright. The 2PIGG catalogue has been produced by application of a percolation algorithm to the 2dFGRS. Resampling the 2PIGG groups with more than 5 confirmed members at $0.076 < z < 0.16$ according to the ratio of overall sample sizes yields a set of about 230 groups with the range of velocity dispersions and number of confirmed members shown in Fig. 15, assuming no cosmic evolution in group properties and a sampling × success rate product of 0.8 relative to 2dFGRS. Thus zCOSMOS-bright should produce an impressive set of well-defined groups in the $10^{13}$-$10^{15}$ $M_\odot$ range for comparison with local 2PIGG equivalents.

### 7.2 Comparison with the VIMOS VLT Deep Survey

The VIMOS VLT Deep Survey is a large spectroscopic redshift survey of, when completed, of order $10^5$ galaxies from three purely magnitude selected samples: the VVDS-Wide going down to $I_{AB} = 22.5$, the VVDS-Deep to $I_{AB} = 24$, and the VVDS-Very Deep down to $I_{AB} = 24.75$ (Le Fèvre et al. 2005). The VVDS-Wide has exactly the same magnitude limit as the COSMOS_bright sample and therefore covers the same redshift range $0.1 < z < 1.4$. The VVDS-Wide covers 10 degree$^2$ in four fields, from which cosmic variance can be estimated on scales up to 100 Mpc, an important information relevant for science analysis in the COSMOS field. The zCOSMOS-bright spectra have a higher spectral resolution (R ~ 600 compared to R ~ 230) and a much denser sampling rate, opening up a lot of the structure dynamical science, and also benefits from an HST-based selection.

The VVDS-Deep selects at $I_{AB} = 24$, 1.5 magnitudes fainter than zCOSMOS-bright and comparable in faintness to the $B_{AB} < 25$ of zCOSMOS-deep. The redshift distribution of the magnitude-selected VVDS-Deep peaks at $z \sim 0.75$ with a long tail extending to very high redshifts $z \sim 5$. The bulk of the galaxies at $z < 1$ thus reach significantly deeper into the luminosity function (Ilbert et al. 2005) than zCOSMOS-bright. On the other hand, at higher redshifts, the BzK and BM-BX color selection of galaxies in the zCOSMOS-deep has been chosen to efficiently isolate the high redshift tail in order to assemble a very large sample of many thousand galaxies with $1.5 < z < 2.5$. The straight magnitude-limited selection of VVDS enables us to estimate the effect of the zCOSMOS-deep color selection function.

### 7.3 Comparison with DEEP2

The DEEP2 redshift survey (Davis et al. 2003, Faber et al. 2006) makes an interesting comparison with zCOSMOS-bright. When completed, which will be well before zCOSMOS, it will contain about 40,000 spectroscopic redshifts in four widely separated fields covering 3.5 deg$^2$. DEEP2 galaxies have $R_{AB} < 24.1$ and are colour selected to lie between $0.7 < z <$ 1.4, more or less straddling the range between zCOSMOS-bright, which has a broad peak around $z \sim 0.7$ and zCOSMOS-deep $1.4 < z < 3.0$. Relative to zCOSMOS-bright, DEEP2 is larger, and has a higher median redshift $<z> \sim 0.91$ vs. 0.65. The DEEP2 survey also has higher resolution spectra and better velocity accuracy (although this is not required for the science investigations of large scale structure and group science).

On the other hand, the product of the sampling rate times the redshift success rate should be higher for zCOSMOS, and once the higher redshifts are accounted for, zCOSMOS actually reaches further down the luminosity function (e.g. $M_{B,AB} < -19.8$ at $z = 0.7$ as against $M_{B,AB} < -20.6$ at $z = 1$ for DEEP2). The combination of higher sampling rate and (effectively) deeper target sample may be important for the definition and characterization of lower mass groups. Finally, DEEP2 is selected off ground-based CFHT images in the rest-frame ultraviolet (3300 Å at $z = 1$), whereas zCOSMOS-bright is selected from the HST/ACS COSMOS images at a rest wavelength comfortably above the 4000 Å break region (5000 Å at z = 0.7), facilitating quantitative comparisons with locally selected samples of galaxies from the SDSS and 2dFGRS. The zCOSMOS-deep survey will of course extend to twice the maximum redshifts of DEEP2 and will be a rather different kind of survey.

## 8. FUTURE SCHEDULE AND DATA RELEASE PRODUCTS

zCOSMOS observations use almost all of the available Service Mode dark observation time on the VLT/UT3. Nevertheless, because it is targeted on a single field, it will still take 4-5 years to complete. By the end of the 2006 observing season, we will have obtained approximately 10,000 bright spectra. Based on the current rate of execution, we project a completion of zCOSMOS-bright observations by July 2008, and of the zCOSMOS-deep observations a year later.

zCOSMOS is undertaken as an ESO Large Program and the team has a contractual obligation to release into the ESO Science Data Archive, at the time of our publication of scientific results, a set of scientific data products from the project. It is anticipated that there will be several phased releases. These will include the input target catalogues, the wavelength- and flux-calibrated spectra, redshifts and Confidence Classes and catalogues of emission line fluxes and equivalent widths, with uncertainties, as well as measurements of other spectral features such as D4000 and selected absorption lines, e.g. Hδ, together with other information (e.g. on target sampling etc) that may be required to utilize the data.

## 9. SUMMARY

zCOSMOS is a large redshift survey being undertaken on the COSMOS field using 600hr of observing time with the VIMOS spectrograph on the 8-m VLT. When completed it will provide a characterization of structure in this region out to redshifts of $z \sim 3$ enabling full realisation of the COSMOS goals to understand the physical links between the evolution of galaxies, their active nuclei, and their environments on scales from the 100 kpc scales of local group environments up to the 100 Mpc scales of the cosmic web. The spectra from zCOSMOS will also provide unique diagnostics on individual objects, including classification, star-formation rates, reddening, metallicities and stellar population diagnostics. zCOSMOS is divided into two parts:

zCOSMOS-bright is a relatively bright sample of 20,000 $I_{AB} < 22.5$ galaxies ($0.1 < z < 1.2$) which span the whole COSMOS field with a high and uniform sampling rate of about 70% and a high success rate in securing redshifts, up to 97% in the main $0.5 < z < 1.0$ range. With a velocity accuracy of well below 100 $kms^{-1}$, this sample is designed to be directly comparable at $0.5 < z < 1.0$ to the 2dFGRS at $z \sim 0.1$ enabling detailed comparisons of galaxies, and their group environments. to be made over the last half of the Hubble time. We already have enough data to establish the reliability and velocity accuracy of the redshifts and to make detailed comparisons with photometrically estimated redshifts, demonstrating the ability of the latter to identify "single-emission-line" spectra and to determine which of the less secure redshifts are most likely to be correct.

zCOSMOS-deep is a fainter survey of approximately 10,000 galaxies that are isolated through colour-selection criteria to have $1.4 < z < 3.0$ in the central 1 $deg^2$ of the COSMOS field. Although our data analysis is less mature for this sample, we have already demonstrated the ability of our selection criteria to isolate galaxies at $1.4 < z < 3.0$ and of our spectra to determine absorption line redshifts for these high redshift galaxies. The success rate approaches that of zCOSMOS-bright for selected subsets of the sample, and especially for targets at $z > 1.9$.

## ACKNOWLEDGEMENTS

Efficient execution of such a large program in Service Mode on the VLT requires the efforts of many individual observatory staff at ESO, both in Garching and on Cerro Paranal, and we gratefully acknowledge the contribution of these many and regrettably sometimes anonymous individuals to our project. The zCOSMOS program builds on many of the hardware and software tools built by the VIMOS/VVDS team, and we gratefully acknowledge here the contribution of those in the VIMOS/VVDS team who are not part of the zCOSMOS project. Finally, we acknowledge with appreciation the contributions of those individuals not listed as authors who have worked, directly or indirectly, to produce the superb imaging data on which the COSMOS survey is based.

TABLES

*Table 1 The zCOSMOS field*

| | center (2000) | | total field size | full sampling region |
|---|---|---|---|---|
| | RA | dec | (RA × dec) | (RA × dec) |
| zCOSMOS-bright | 10 00 28 | 02 13 37 | 1.30° × 1.24° | 0.98° × 0.95° |
| zCOSMOS-deep | 10 00 43 | 02 10 23 | 0.92° × 0.91° | 0.60° × 0.62° |

*Table 2 Preliminary verification of confidence classes in zCOSMOS-bright*

| | Fraction of sample | Spectroscopic verification rate[a] | Photometric consistency rate ($\chi^2 < 1.5$)[d] | Photometric consistency rate (all $\chi^2$)[e] |
|---|---|---|---|---|
| Class 4 | 50% | 111/111 = 100% | 510/517 = 99% | 555/567 = 98% |
| Class 3 | 23% | 34/34 = 100% | 220/223 = 99% | 261/268 = 97% |
| Class 9 | 6% | 10/12 = 83%[b]<br>12/12 = 100%[c] | 56/65 = 86%<br>63/65 = 97% | 65/77 = 84%<br>72/77 = 94% |
| Class 2 | 7% | 10/12 = 83% | 59/67 = 88% | 75/84 = 89% |
| Class 1 | 7% | 2/4 = 50% | 36/67 = 54% | 42/79 = 53% |
| Class 0 | 7% | - | - | - |

*Notes to Table 2*

*a* *Computed in terms of the number of "valid tests" from the repeat observations of one mask, see text for details.*

*b.* *Assuming the nominally assigned redshift*

*c.* *Allowing for the alternative redshift inherent in Class 9*

*d.* *Considering only objects with a good template match and regarding consistency as* $|z_s - z_p| < 0.1(1+z)$

*e.* *Considering all objects regardless of the quality of the template match.*

FIGURE CAPTIONS

*Fig 1:* *(a) Simulation of structure revealed by a redshift survey of the characteristics of zCOSMOS-bright (see text for details) as generated from the mock catalogues of Kitzbichler et al (private communication). Each panel (left to right, top to bottom) shows an increment of 0.003 in redshift starting at z = 0.700 (top right). Solid black dots show objects with spectroscopically determined redshifts. Larger shaded grey dots show all galaxies with* $I_{AB} < 24$ *so as to show the underlying structure. The depth of each panel in redshift space is approximately five times the redshift accuracy of the spectroscopic redshifts.*

*(b) The same simulation as Fig 1a showing galaxies inferred to lie in the same volume (0.700 < z < 0.736) as defined with their photometric redshifts, which are assumed to have a* $1\sigma$ *uncertainty of ±0.03(1+z). In this case the depth of this single panel is only 30% of the FWHM of the photometric redshift distribution leading not only to a smoothing of the structure in Fig1a but also the inclusion of galaxies that in reality lie outside of this redshift range.*

*Fig 2:* *The pointing centers and total coverage of the two components of zCOSMOS compared with the mosaic of 600 COSMOS ACS images. The zCOSMOS survey utilizes 90 pointings (with field centers indicated by the small circles) that are uniformly spaced in RA and declination so as to provide a constant sampling rate over a large central contiguous area, with four edge regions with approximately half the central sampling rate and four corner regions with only a quarter of the central sampling rate, as indicated by the regions bounded by straight lines in the figure. zCOSMOS-bright uses all 90 pointings with two mask designs at each position. This provides coverage over almost the entire ACS mosaic and gives each target galaxy in the central region eight opportunities to be observed. zCOSMOS-deep uses a mosaic of 42 central pointings (indicated by crosses) with a correspondingly smaller coverage. The zCOSMOS pointings in the mosaic are designated in (x,y) starting at the lower right. The VIMOS quadrant pattern is shown shaded for pointing (6,5).*

*Fig 3:* *Differential I-band number counts for the zCOSMOS-bright sample for galaxies (solid circles) and stars (open circles). The number counts of galaxies show excellent agreement with those of Postman et al (1998).*

*Fig 4* *Expected sampling as a function of separation in the zCOSMOS-bright sample, based on an analysis of the small region for which all masks have now been designed. The sampling is defined as the fraction of galaxies that are separated by a given amount on the sky from an object that is observed spectroscopically, that are also observed spectroscopically. Because of the multiple passes, giving each object eight opportunities to be included in the mask designs, there is only a weak bias against near neighbours below 10 arcsec separation and none at larger separations. The horizontal dashed line shows the program design goal of a uniform 70% sampling rate. The sampling of the zCOSMOS-deep sample is expected to be similar.*

*Fig 5* *Composite spectra from zCOSMOS-bright showing an early type spectrum and an emission line spectrum.*

*Fig 6* *Redshift differences for the 90 objects in the repeat observations of a single mask. The dotted lines show the redshift difference corresponding to a velocity difference of ± 100* $kms^{-1}$*. The inset shows the distribution of velocity errors which is well represented by a Gaussian of width* $\sigma = 78$ $kms^{-1}$ *implying an uncertainty per measurement of 55* $kms^{-1}$*.*

*Fig 7* *Plot of spectroscopic redshift against photometric redshift for each spectroscopic Confidence Class. Solid symbols represent galaxies with good template fits ($\chi^2$ < 1.5) and open symbols those with poorer fits. For the highly reliable Class 3 and 4 objects, 98% lie within the consistency region with $|z_s$-$z_p| < 0.1\times(1+z)$ and the majority of the discrepancies have in any case poor photometric matches. As the Confidence Class degrades, the fraction of outliers increases as does the fraction of outliers with good template matches, while still maintaining a significant $z_s \sim z_p$ component. There is in fact a very good agreement between the fraction of consistent redshifts and the purely spectroscopic verification rate from the repeat observations (see Table 2), suggesting that the photometric redshifts can be used to indicate which of the redshifts in the less reliable Classes 1 and 2 are likely to be correct and which are likely to be incorrect, as discussed in the text. This is especially true for the Class 9 objects where most of the outliers become consistent at the "alternative" redshift implied by swapping the identification of the single emission line between H$\alpha$ 6563 and [OII] 3727.*

*Fig 8:* *The distribution in color-redshift space of the objects for which a redshift was successfully obtained (left) and those for which a redshift was not obtained (right), plotting the latter at their photometric redshift. Objects in Class 2 and 1 with a consistent photometric redshift defined to be $|z_s$-$z_p| < 0.1\times(1+z)$ as in Fig. 7 are counted as successes and those with an inconsistent photometric redshift, plus the Class 0 objects, are counted as failures.*

*Fig 9:* *The success rate in zCOSMOS-bright as a function of redshift, assuming the photometric redshift for those objects for which no spectroscopic redshift was secure. This shows a broad maximum close to unity at z ~0.7, which is also the peak redshift for this $I_{AB}$ < 22.5 sample. The success rate falls towards about 80% both at high (z > 1) and very low redshifts (z < 0.3). The average for the sample as a whole is currently 89%.*

*Fig 10* *Histogram of redshifts of galaxies in the first set of zCOSMOS-bright spectra, showing strong structure in redshift space. The upper panel shows the redshifts of objects with less secure spectroscopic redshifts in Classes 2 and 1 whose redshifts are "confirmed" with photometric redshifts – see text for details. Redshift bins have $\Delta z$ = 0.001.*

*Fig 11* *The distribution in Right Ascension and declination of galaxies observed so far in zCOSMOS-bright (see Fig 10). These objects cover about one fifth of the COSMOS field. The visible banding structure in RA and declination reflects the currently highly variable coverage due to the quadrant design of VIMOS. The final surveys will have uniform coverage across a large central region (see Fig 2).*

*Fig 12* *Composite spectra from zCOSMOS-deep, showing galaxies with Lyman $\alpha$ in emission and in absorption. Both spectra show a host of absorption features.*

*Fig 13* *Distribution in redshifts of the current set of 977 zCOSMOS-deep spectra. There is a small contamination from stars and a larger contamination from very low redshift galaxies. The upper left panel shows the redshift distribution of all objects that were actually observed in 2005 with Confidence Classes 3 and 4 (black), 2 and 9 (hatched) and Class 1 (open). Objects with no redshift (Class 0) are represented by the detached rectangular area. The remaining panels show the same information for those subsets of these objects that, with our improved G and K photometry, now satisfy the BX, BM and BzK criteria. The horizontal bars show the expected range*

*of redshifts. It is noticeable how the BX targets have a higher success rate than the lower redshift BM targets, with the BZK in between, reflecting their broader N(z). In order to match the redshift distribution of the BM objects with that of Steidel et al (2004) we would need that most of the failures (Class 0), and most of those unreliable redshifts (Class 1) which will be found to be incorrect, lie at $z < 2$.*

*Fig 14* *Comparison of the absolute magnitudes of galaxies in the low redshift 2dFGRS (black symbols, upper scale) with the first galaxies observed in zCOSMOS-bright (grey symbols, lower scale), removing an assumed luminosity evolution of $\Delta M_B = z$ magnitudes from the latter. The zCOSMOS-bright galaxies at $0.5 < z < 1.0$ are well-matched both in rest-frame selection wavelength and in luminosity, and as discussed in the text, broadly similar in the product of sampling and success rate, as well as in velocity accuracy. The final zCOSMOS-bright will provide an ideal comparison sample for the 2dFGRS at a look-back time of a half the Hubble time, albeit only one tenth the size.*

*Fig 15* *Simulation of a zCOSMOS-derived group catalogue extracted from the 2dFGRS 2PIGG group catalogue of Eke et al between $0.076 < z < 0.16$, see Fig 11 and text for details. A relative sampling $\times$ success rate of 80% between the samples is assumed.*

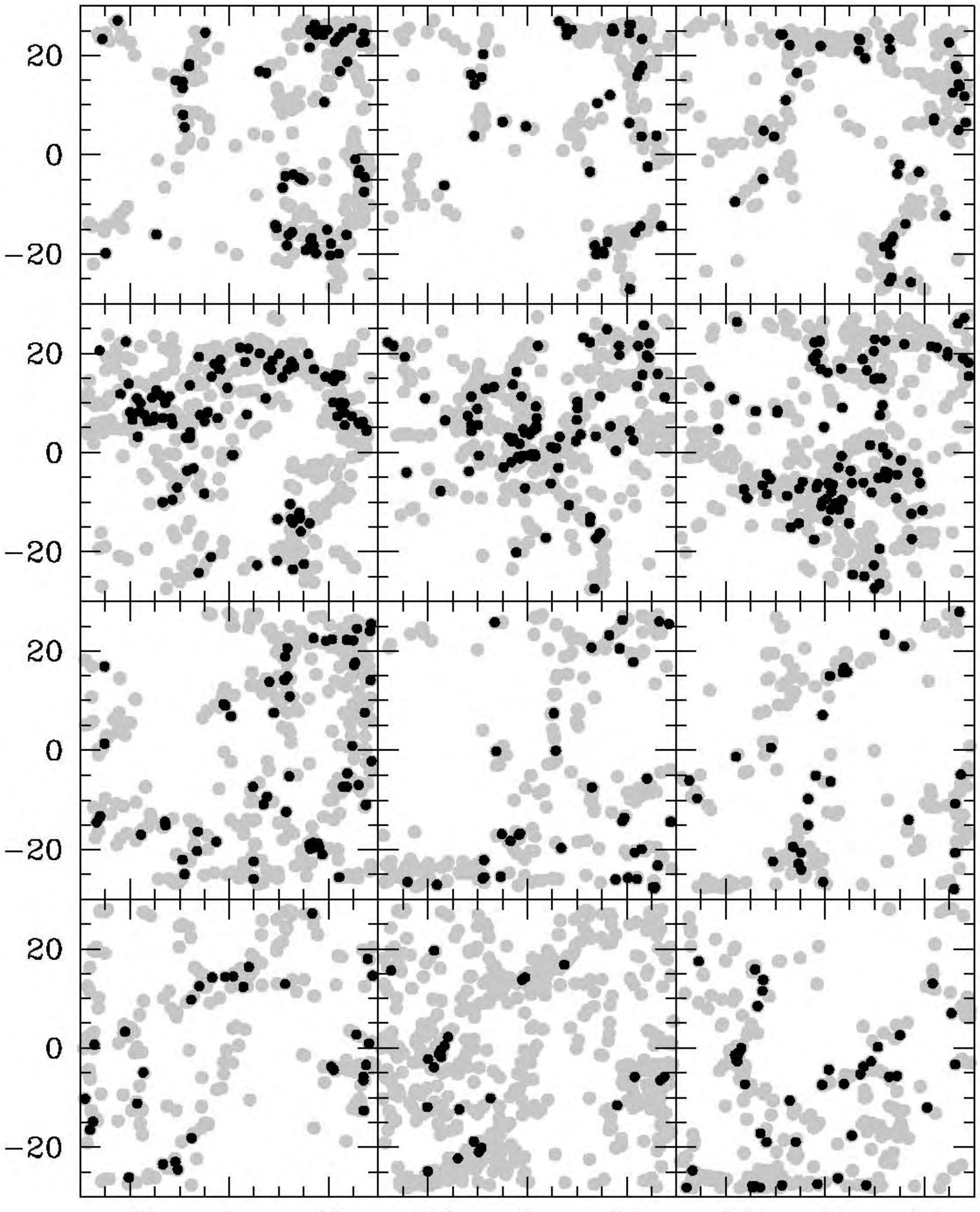

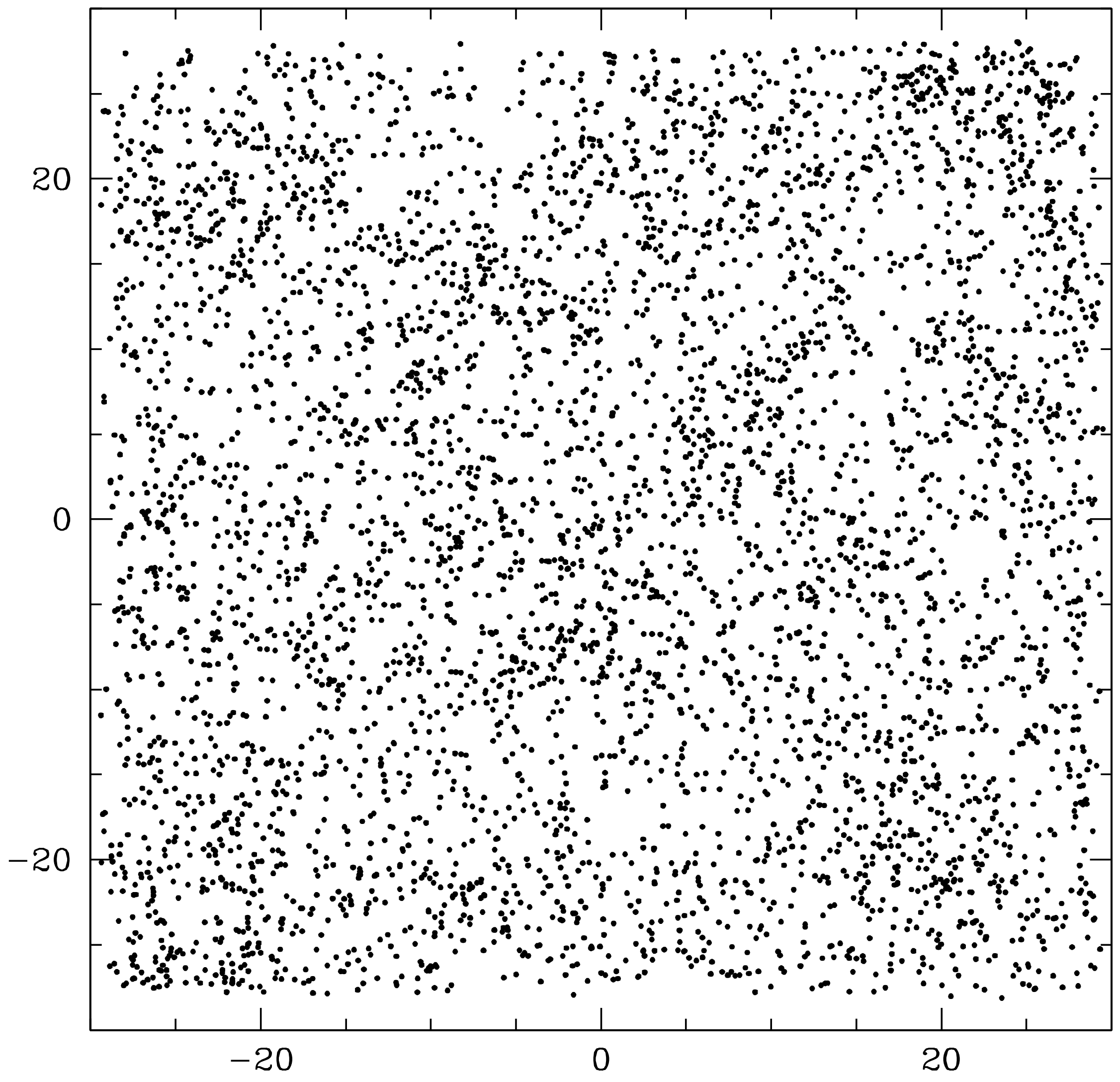

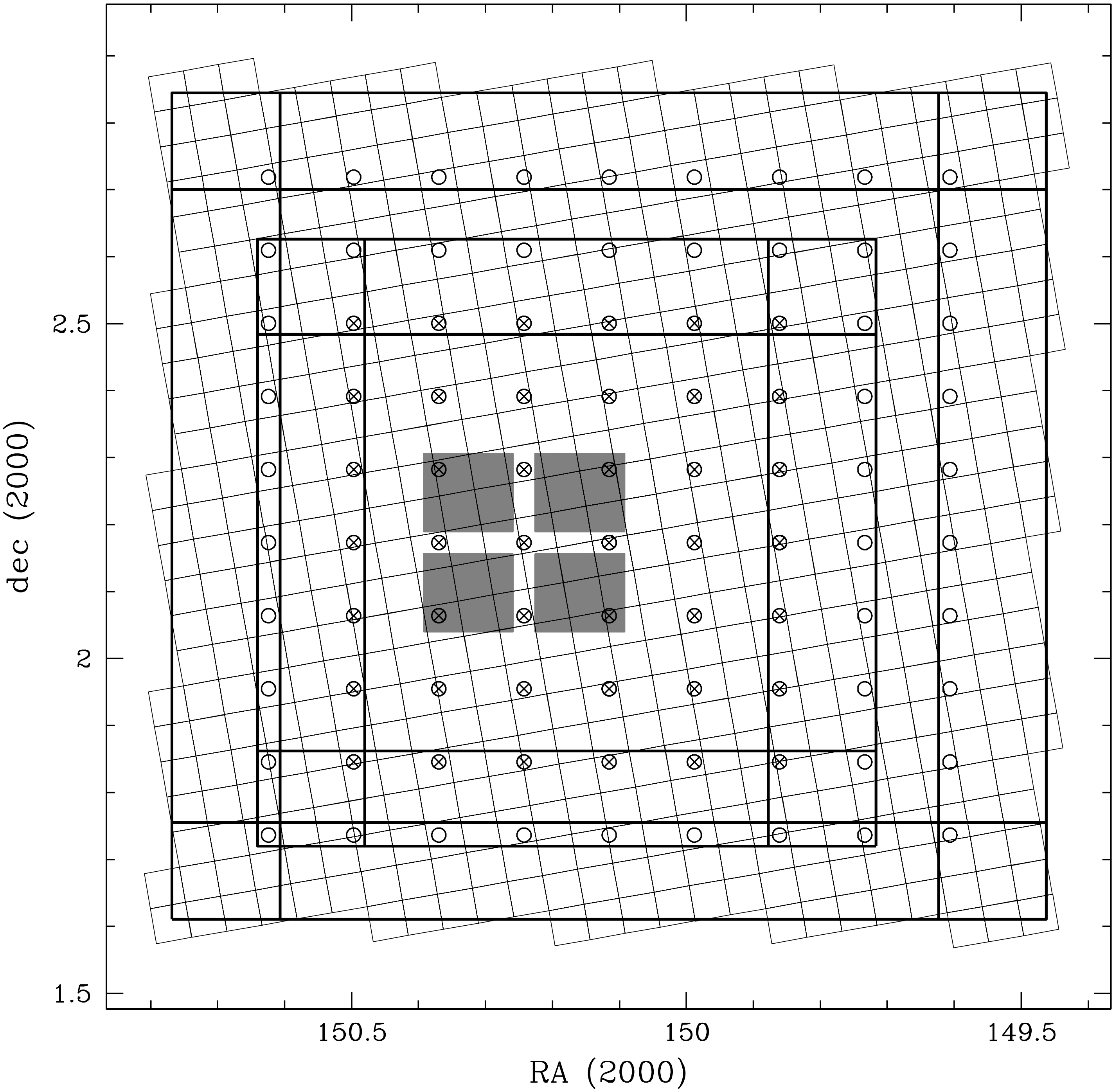

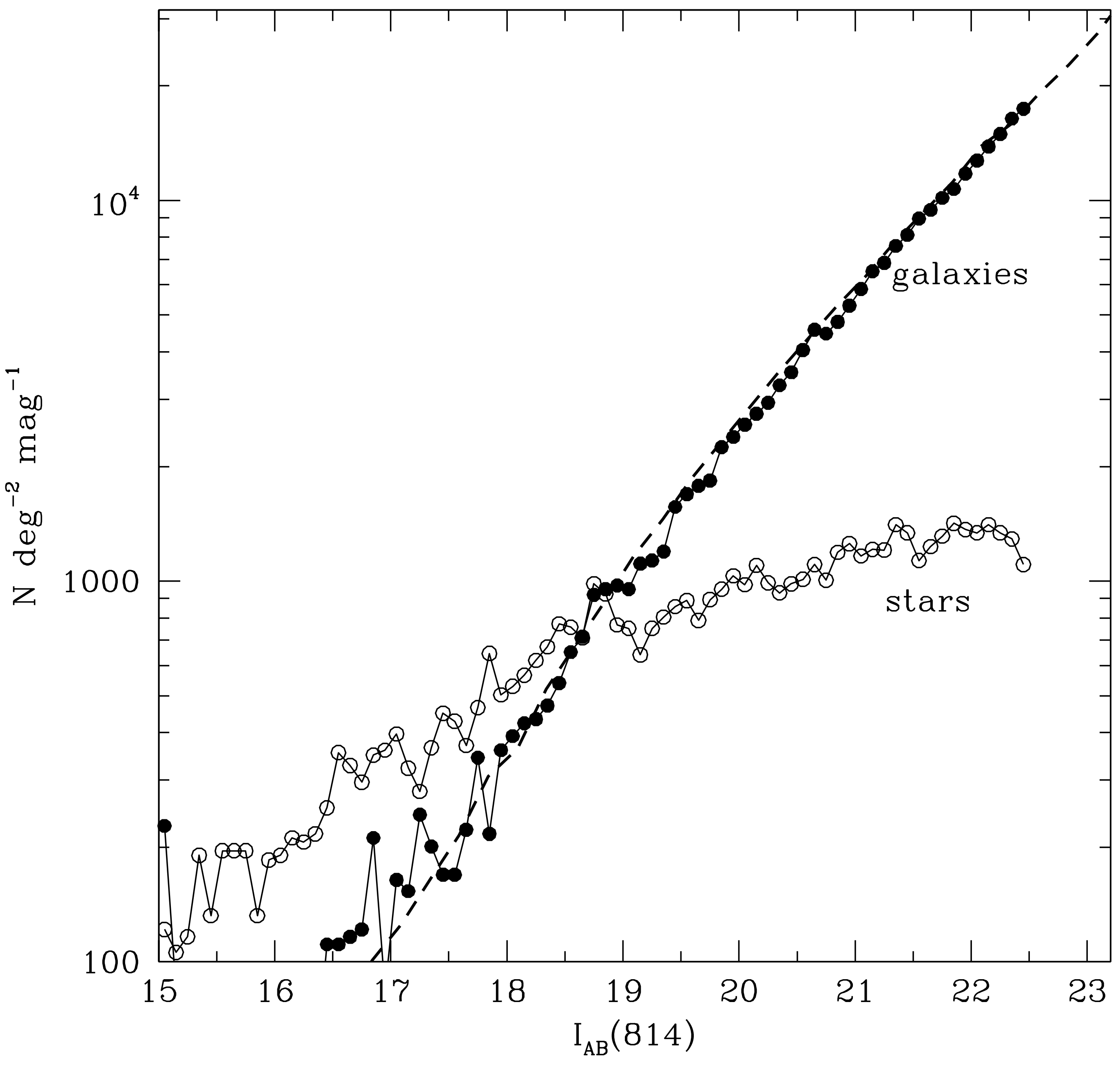

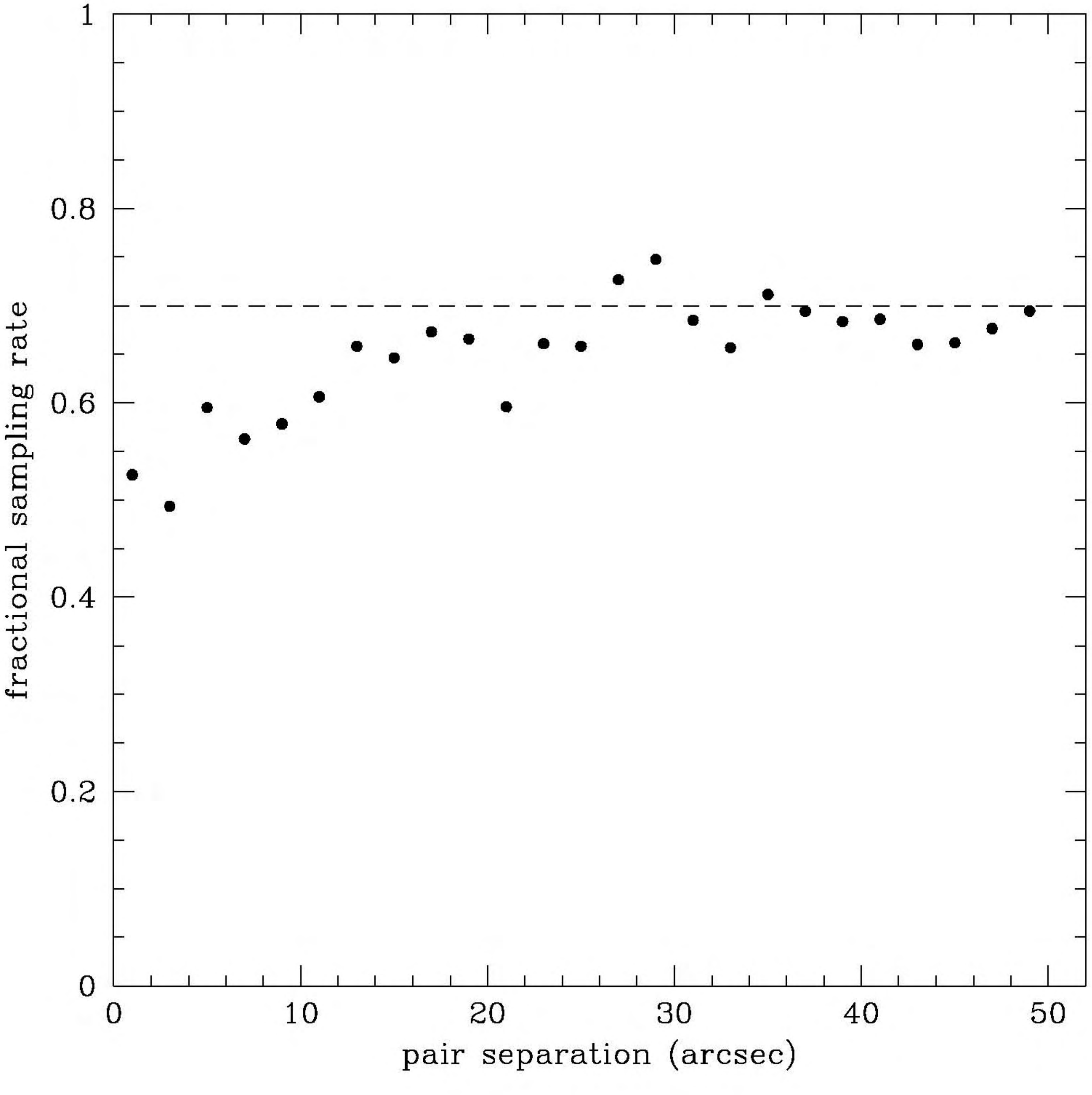

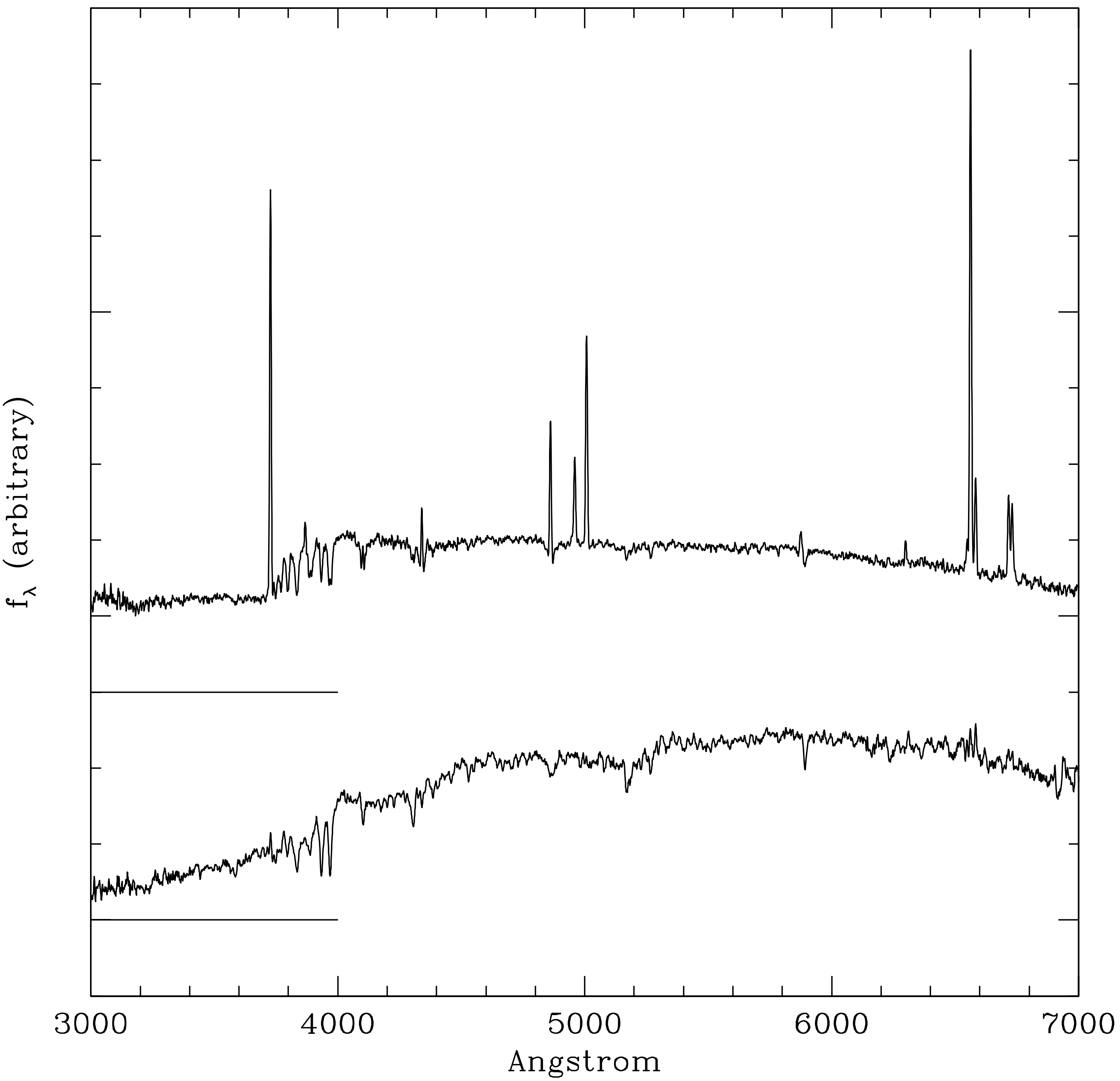

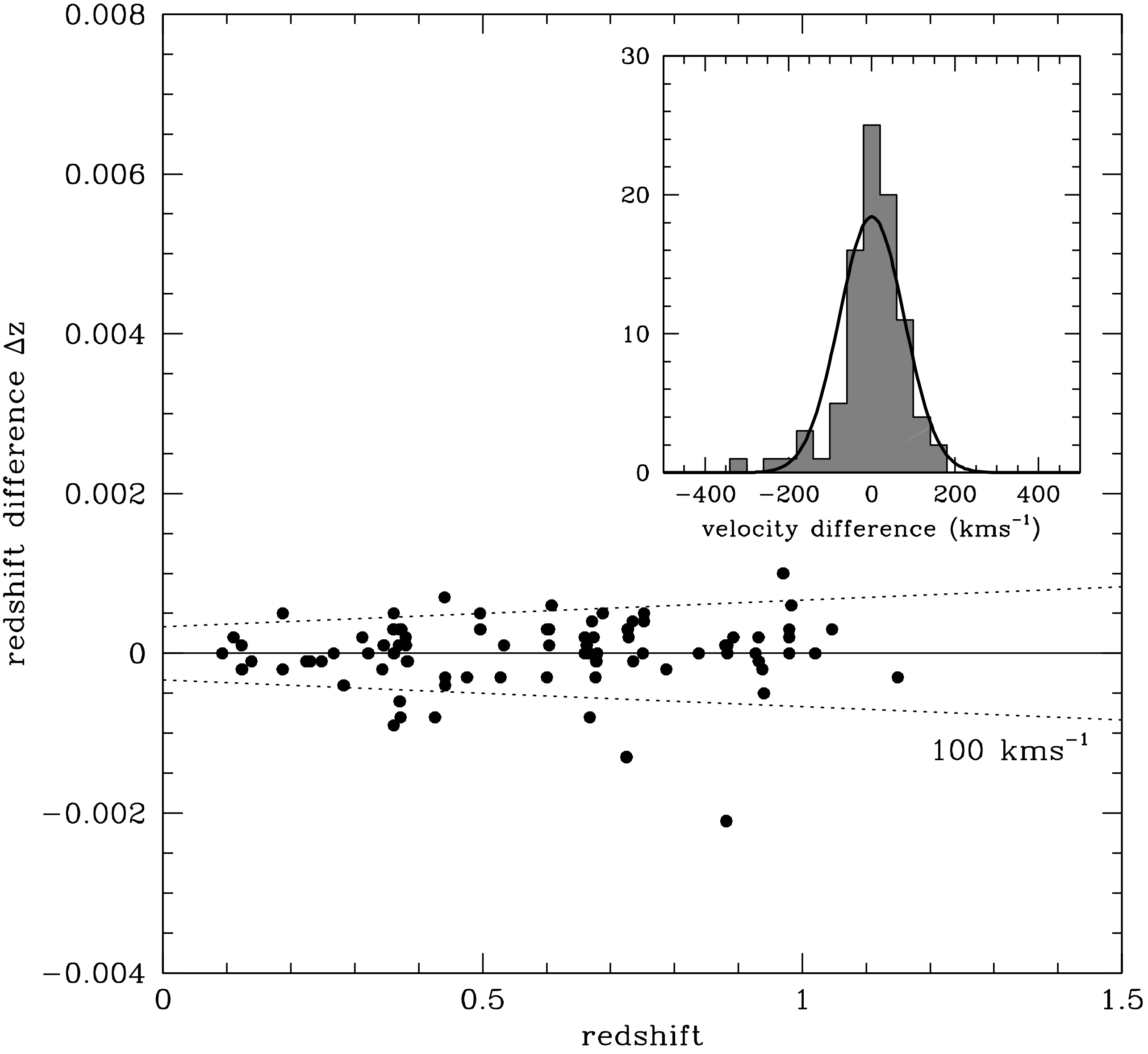

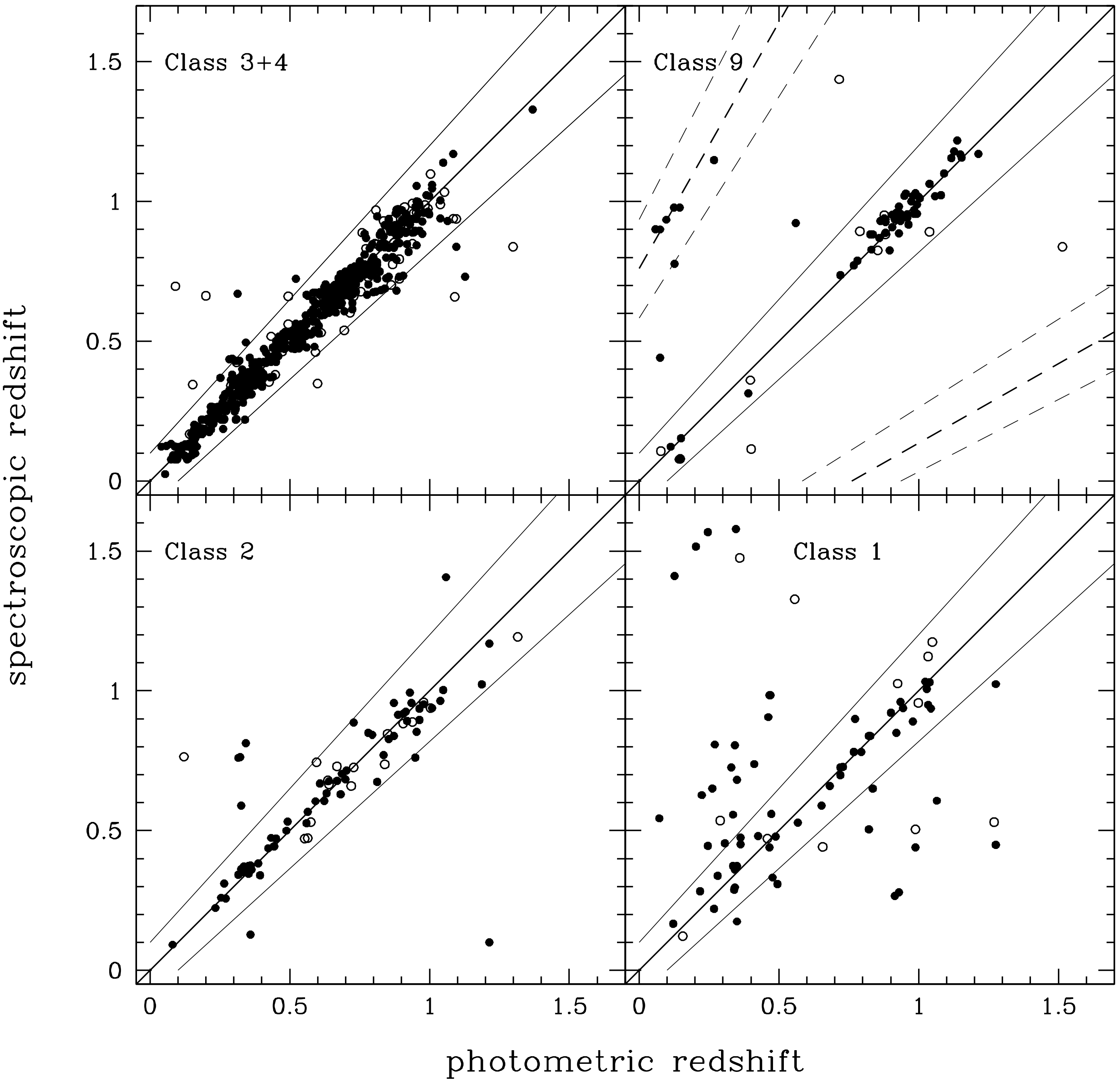

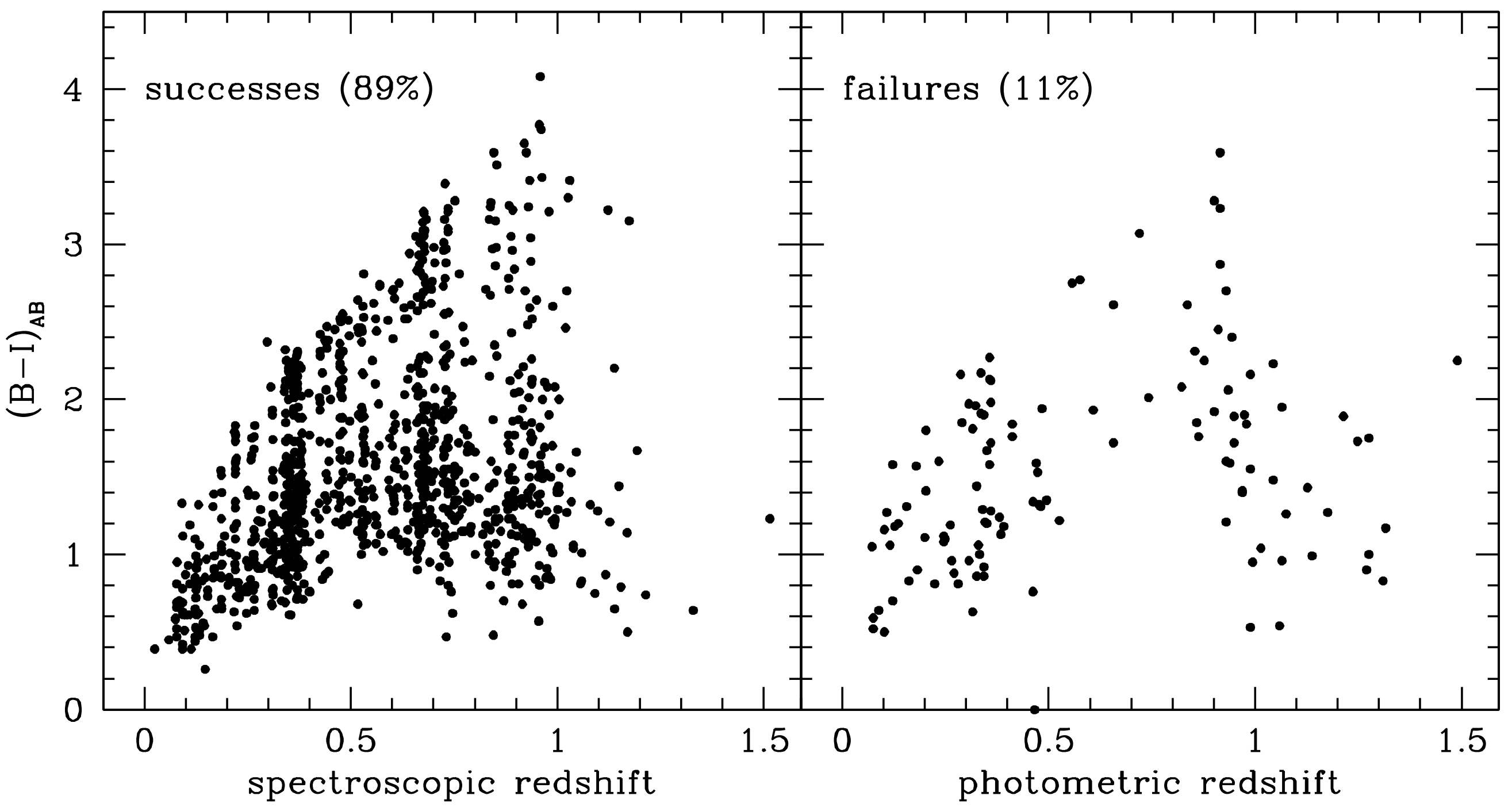

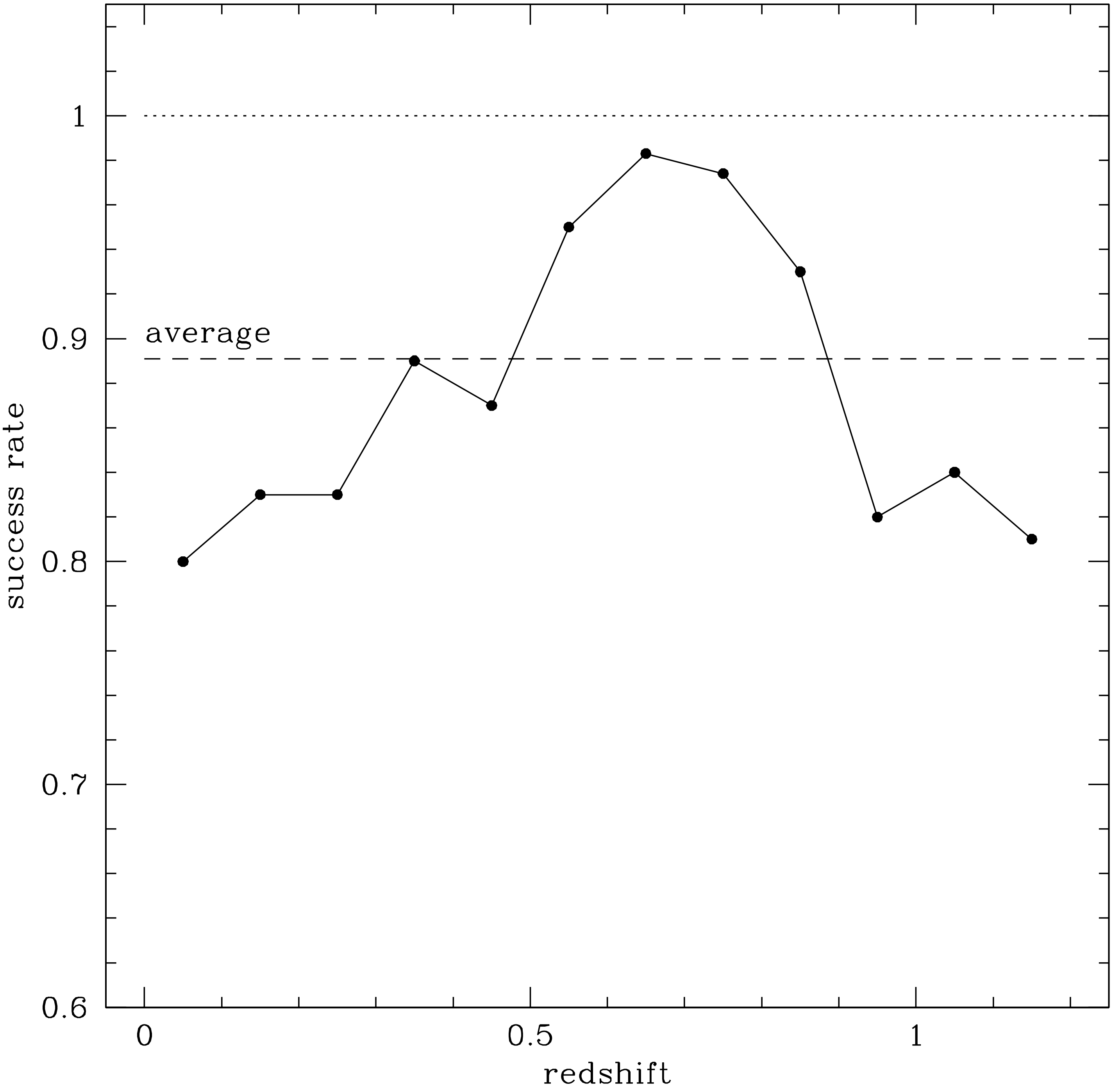

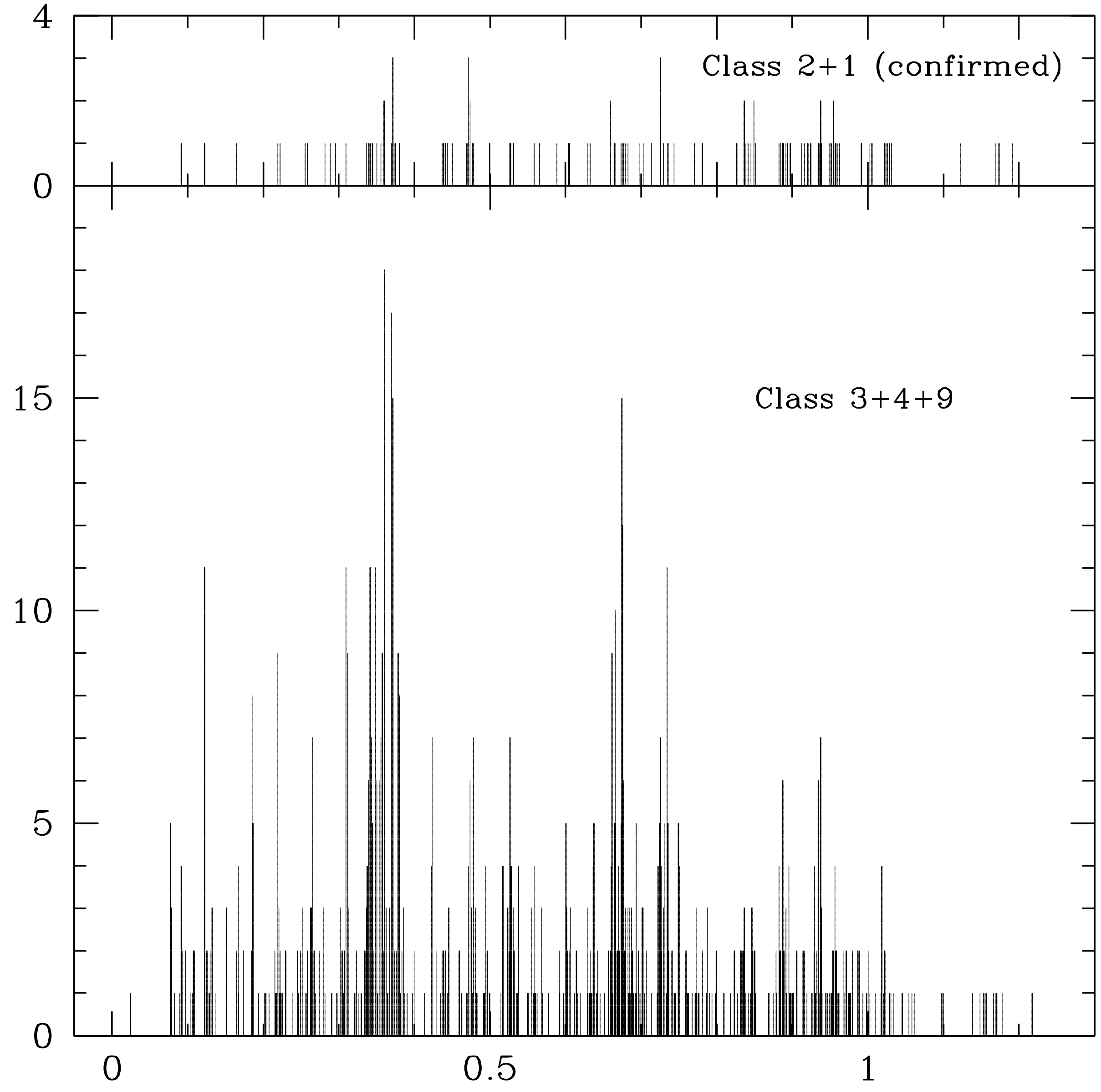

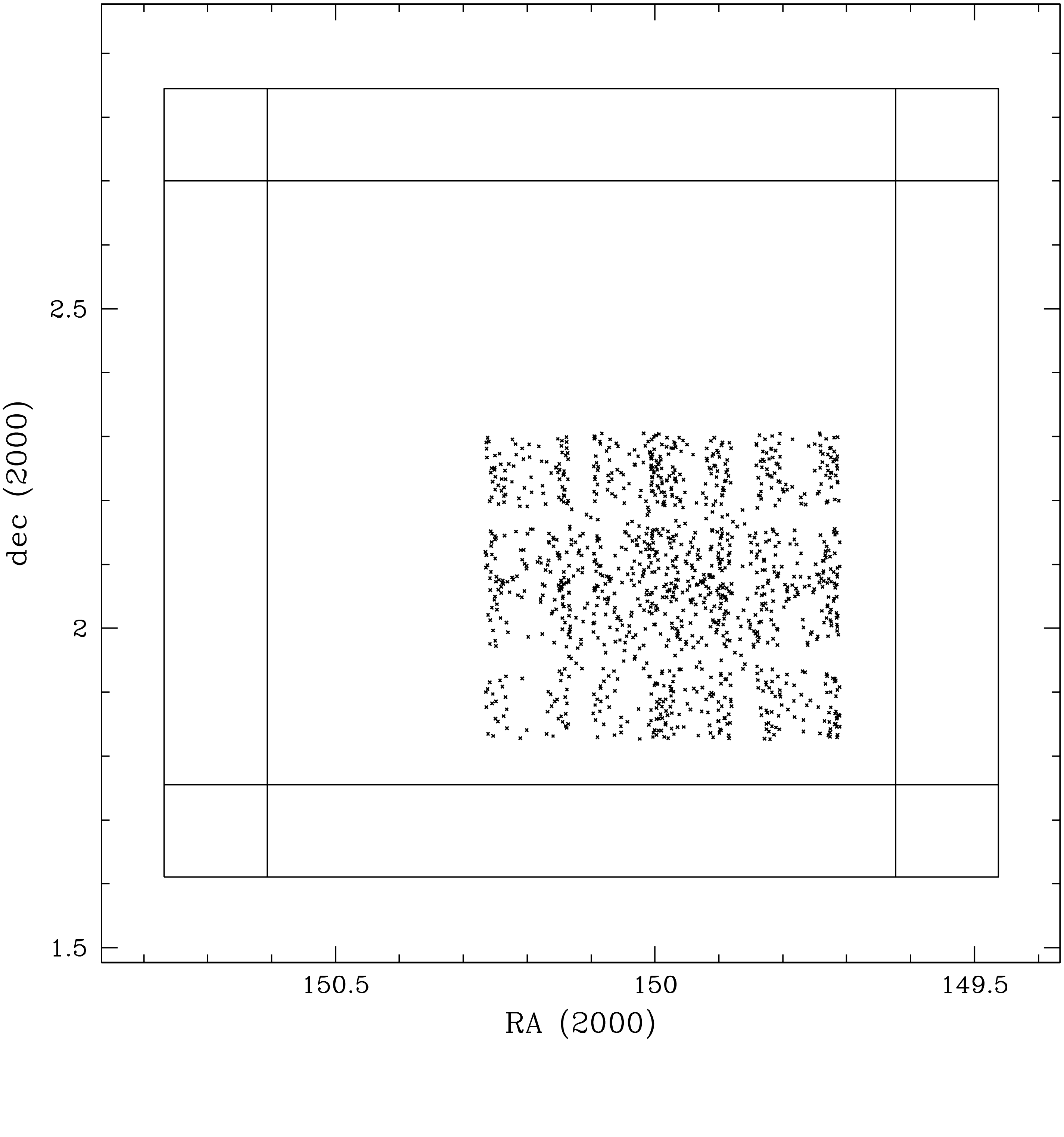

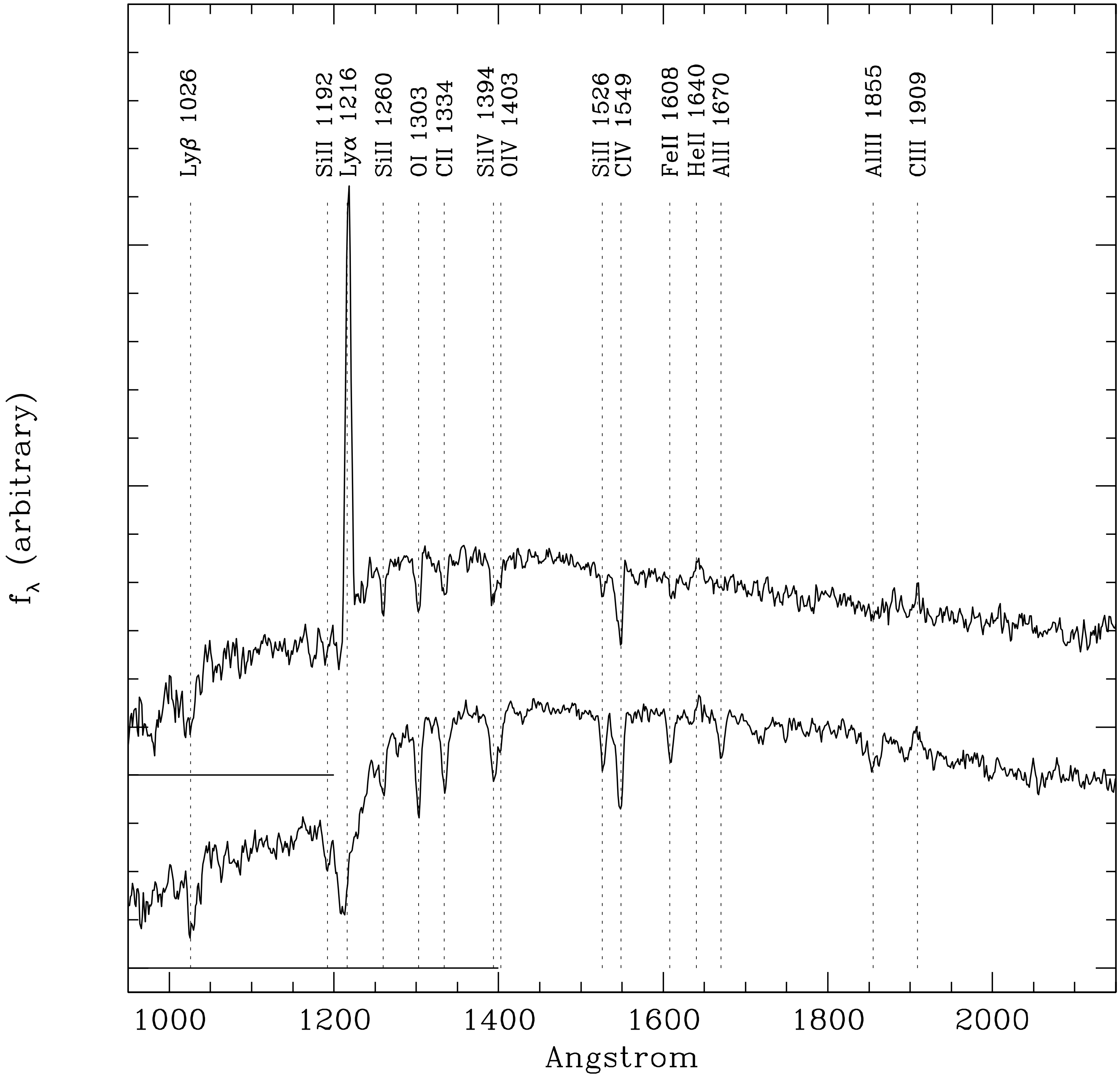

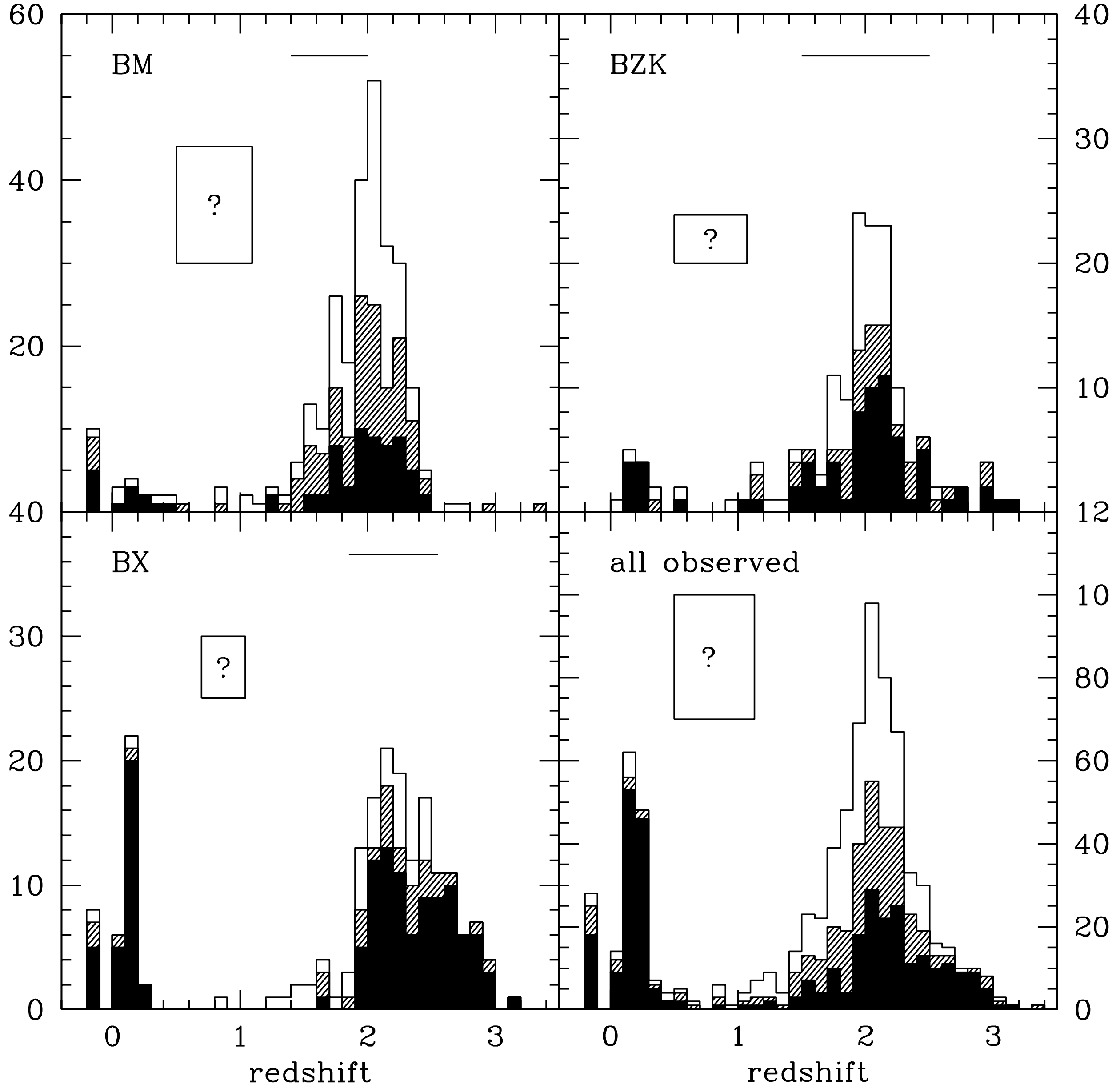

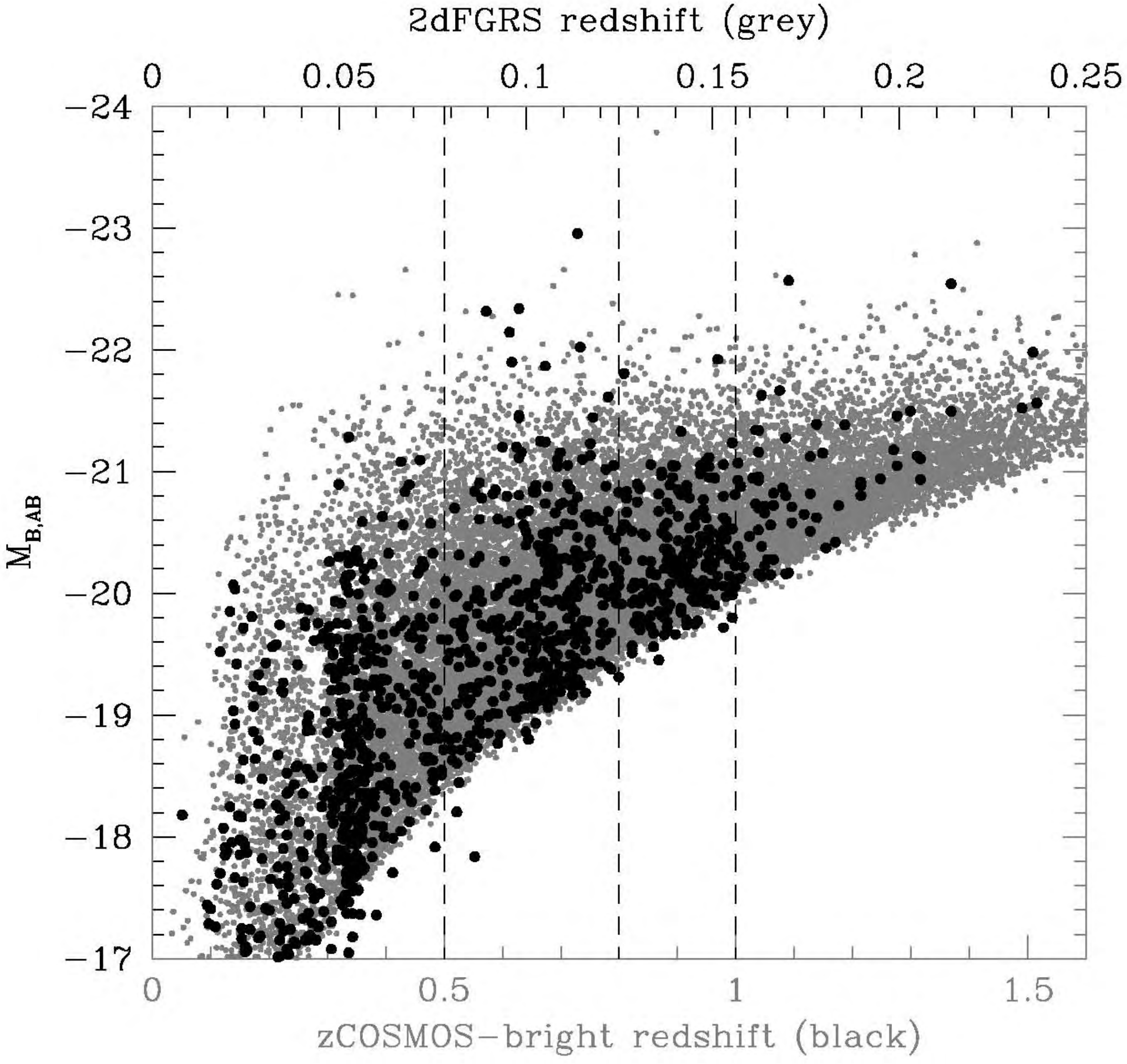

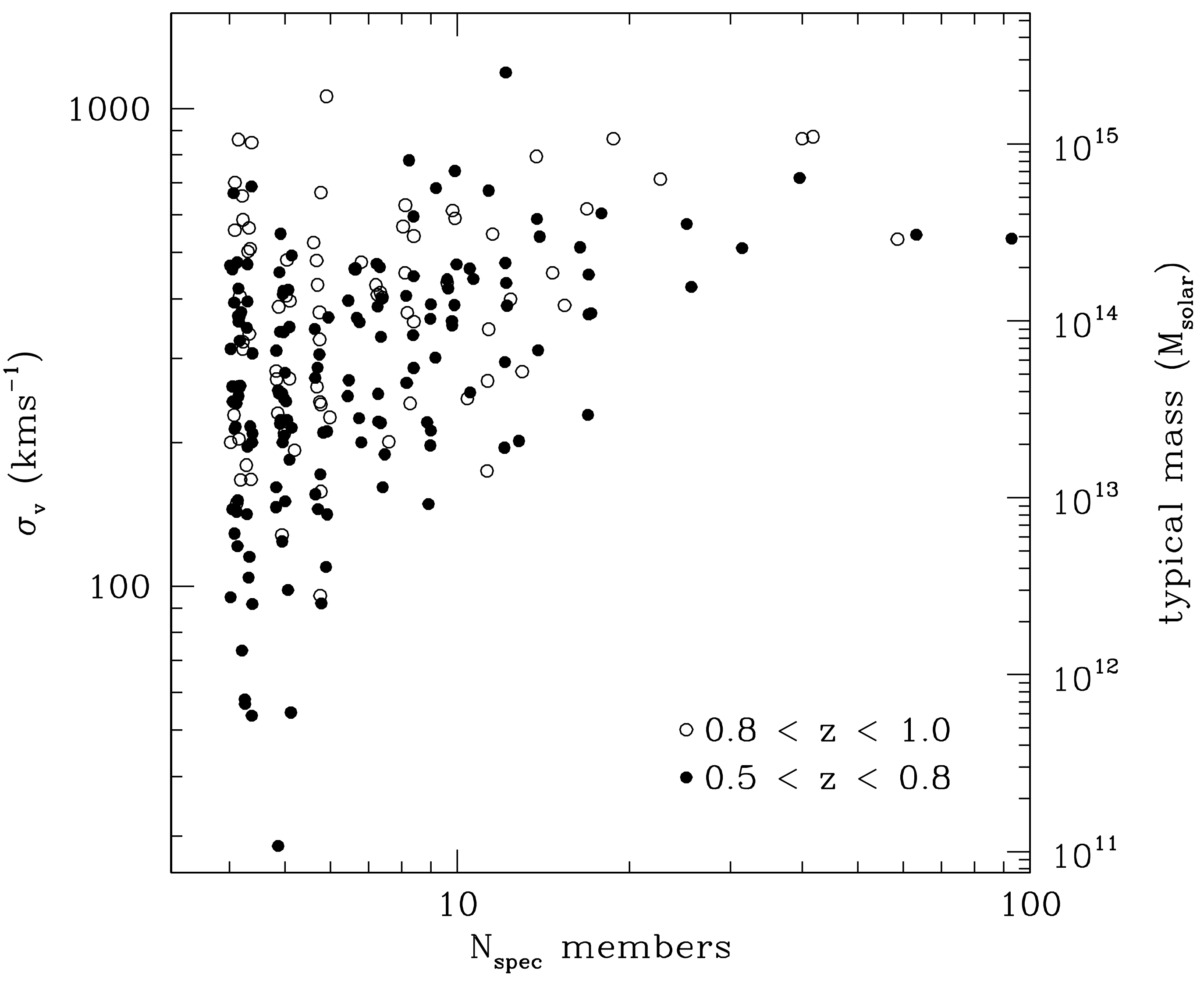